\documentclass[12pt,a4paper,oneside,pdftex]{article}

\usepackage[english]{babel}
\usepackage[utf8]{inputenc}
\usepackage[T1]{fontenc}

\usepackage{lmodern}

\usepackage{latexsym}

\usepackage{geometry}

\newtheorem{EXAMPLE}{Example}[section]
\newtheorem{theorem}{Theorem}[section]

\newtheorem{lemma}[theorem]{Lemma}
\newtheorem{proposition}[theorem]{Proposition}

\newtheorem{REMREM}{Remark}[section]
\newtheorem{DEFINITION}{Definition}[section]

\newenvironment{example}{\begin{EXAMPLE}\small}{\end{EXAMPLE}}

\newenvironment{definition}{\begin{DEFINITION}\rm}{\end{DEFINITION}}

\newcommand{\VAR}{\mbox{$Var$}}
\newcommand{\TRUE}{\mbox{$True$}}
\newcommand{\FALSE}{\mbox{$False$}}
\newcommand{\FUNC}[1]{\mbox{$Func_{#1}$}}
\newcommand{\PRED}[1]{\mbox{$Pred_{#1}$}}
\newcommand{\TERM}[1]{\mbox{$Term_{#1}$}}

\newcommand{\FEA}[1]{\mbox{$EQ_{#1}$}}

\newcommand{\EQNB}[1]{\mbox{$Eqn_{#1}/_{\approx}$}}

\newcommand{\SUBEMPTYA}{\mbox{$\varepsilon$}}
\newcommand{\SUBEMPTYB}{\mbox{$\varepsilon_{\approx}$}}
\newcommand{\SUBFALSE}{\mbox{$\bot$}}
\newcommand{\SUBA}[1]{\mbox{$Sub_{#1}$}}
\newcommand{\SUBB}[1]{\mbox{$Sub_{#1}^{\bot}$}}
\newcommand{\SUBC}[1]{\mbox{$Sub_{#1}^{\bot}/{\approx}$}}

\newcommand{\TRUTH}[3]{\mbox{\( #1 \models_{#2} #3 \)}}
\newcommand{\NOTTRUTH}[3]{\mbox{\( #1 \not \models_{#2} #3 \)}}
\newcommand{\RULEA}[3]{\mbox{\( #1 \stackrel{#2}{\Longrightarrow} #3 \)}}

\newcommand{\REWRA}[3]{\mbox{\( #1 \stackrel{#2}{\longrightarrow} #3 \)}}
\newcommand{\REWRB}[2]{\mbox{\( #1 \longrightarrow^{\star} #2 \)}}

\newcommand{\SOLN}[2]{\mbox{$soln_{#1}(#2)$}}

\newcommand{\ANS}[3]{\mbox{$\langle #1 \rangle_{#2}^{#3}$}}

\newcommand{\BB}[2]{\mbox{$B_{#1}^{#2}$}}
\newcommand{\HA}{\mbox{$H$}}
\newcommand{\HB}{H}
\newcommand{\UU}[2]{\mbox{$U_{#1}^{#2}$}}
\newcommand{\VV}[2]{\mbox{$V_{#1}^{#2}$}}

\newcommand{\DOM}[1]{\mbox{$dom(#1)$}}
\newcommand{\RANGE}[1]{\mbox{$range(#1)$}}
\newcommand{\VARS}[1]{\mbox{$vars(#1)$}}
\newcommand{\ELIM}[1]{\mbox{$elim(#1)$}}
\newcommand{\PARAM}[1]{\mbox{$param(#1)$}}
\newcommand{\BOUND}[1]{\mbox{$bound(#1)$}}
\newcommand{\KERNEL}[1]{\mbox{$kernel(#1)$}}
\newcommand{\RESTR}[2]{\mbox{$#1 | #2$}}

\newcommand{\EQ}[1]{\mbox{${\cal E}$}-#1}
\newcommand{\BINDA}[2]{\mbox{$#1 \leftarrow #2$}}

\newcommand{\CARD}[1]{\mbox{$|#1|$}}
\newcommand{\CLAUSE}[2]{\mbox{$#1 \leftarrow #2$}}
\newcommand{\END}{\mbox{$\Box$}}

\newcommand{\CCB}[3]{#1[#2 \leftarrow #3]}

\newcommand{\CMA}{\mbox{${\cal R}$}} 
\newcommand{\CMB}[1]{\mbox{${\cal #1}$}} 
\newcommand{\IND}[3]{\mbox{$Ind_{#1}(#2,#3)$}} 

\newcommand{\LA}[1]{\mbox{$L_{#1}$}}
\newcommand{\LB}[1]{L_{#1}}
\newcommand{\MM}[2]{\mbox{$M_{#1}^{#2}$}}
\newcommand{\OLA}[2]{\mbox{$\overline{#1}#2$}}
\newcommand{\OLB}[2]{\overline{#1}#2}
\newcommand{\MTRUTH}[3]{\mbox{${#1}\,\mid\!\equiv_{#2}\,{#3}$}}
\newcommand{\MANS}[3]{\mbox{$[#1]_{#2}^{#3}$}}
\newcommand{\EMPTYSEQA}{\mbox{$\Lambda$}}
\newcommand{\EMPTYSEQB}{\Lambda}
\newcommand{\HC}{}
\newcommand{\TOA}[2]{\mbox{$T_{#1}^{#2}$}}
\newcommand{\TOBA}[3]{\mbox{$T_{#1}^{#2}(#3)$}}
\newcommand{\TOBB}[3]{T_{#1}^{#2}(#3)}
\newcommand{\TOCA}[3]{\mbox{$T_{#1}^{#2} \uparrow #3$}}
\newcommand{\TOCB}[3]{T_{#1}^{#2} \uparrow #3}

\newcommand{\RWA}[1]{\mbox{${\cal W}_{#1}$}}
\newcommand{\RWB}[1]{{\cal W}_{#1}}
\newcommand{\SR}[1]{\mbox{${\cal S}_{#1}$}}
\newcommand{\CM}[1]{\mbox{${\cal R}_{#1}$}}

\newcommand{\EMPTYINDA}{\mbox{{\small $\Lambda$}}}

\usepackage[pdftex,bookmarksnumbered,pagebackref]{hyperref} 

\hypersetup{
 pdftitle={Existentially Quantified Systems of Equations as an Implicit Representation of Answers in Logic Programming},
 pdfauthor={Ján Komara},
}

\hypersetup{
 unicode,
 colorlinks=true,
 allcolors=[rgb]{0,0.5,0.5},
 urlcolor=[rgb]{0.58,0.0,0.83},
 breaklinks=true,
}

\begin{document}

\title{
 Existentially Quantified Systems of Equations as an Implicit Representation of Answers in Logic Programming%
 \thanks{Reprint of the technical report TR mff-ii-11-1992, September 1992.}
 \thanks{The shorter version of this paper is titled \emph{Equations or Substitutions in Logic Programming?} and can be found in Procceedings of Logic Programming Winter School and Seminar, Lop'92 :- !., Rusava, Czechoslovakia, 1992.}
}
\author{
 {\Large Komara J\'{a}n} 
 \and
 {\small Institute of Informatics, MFF UK, Mlynsk\'{a} dolina, 842 15 Bratislava, Czechoslovakia}
}

\date{}

\maketitle

\begin{abstract}
In this paper we present an alternative approach to formalize the theory of
logic programming. In this formalization we allow existential quantified
variables and equations in queries. In opposite to standard approaches the role
of answer will be played by existentially quantified systems of equations. This
allows us to avoid problems when we deal with substitutions. In particular, we
need no ''global'' variable separated conditions when new variables are
introduced by input clauses. Moreover, this
formalization can be regarded as a basis for the theory of concurrent logic
languages, since it also includes a wide spectrum of parallel computational
methods. Moreover, the parallel composition of answers can be defined directly
--- as a consistent conjunction of answers.
\end{abstract}

\section{Introduction}\label{sc:intr}

In the theory of logic programming we deal with three kinds of semantics. The
declarative and fixpoint semantics are based on a rigorous mathematical theory.
On the other side, there are serious difficulties in describing the operational
semantics. For example,  standard approaches to SLD-resolution are based on the
explicit representation of answer --- in terms of substitutions. The key issue
is in finding a suitable class of substitutions and a variable separated
condition to ensure completeness (see \cite{KN:revisited}). In this paper we
present an other formalization of the theory of logic programming which is
based on an implicitly representation of answer. Answers are defined in terms
of existentially quantified systems of equations. By allowing of existentially
quantified variables we eliminate problems concerning renaming of input
clauses. We show that there is a strong relationship between the both
representation of answer. We give an algorithm, called Solved Form Algorithm,
which transforms any existentially quantified system of equations into a solved
form from which the corresponding (finite) substitution can be reached
directly.  Finally, this formalization can be regarded as a basis for the
theory of concurrent logic languages, since the parallel composition of answers
can be defined directly --- as a consistent conjunction of answers.

The fundamental notion of logic programming is query answering. Given a logic
program $P$ and a query
\( Q \equiv A_1 \wedge \ldots \wedge A_m \)
with (free) variables
\( \VARS{Q} = \{ x_1, \ldots, x_n \} \)
we find an answer to questions of the form ''for which values of
\( x_1, \ldots, x_n \)
the query $Q$ is true w.r.t. the program $P$ ?''. The standard approach to
query answering may be seen as finding the explicit representation of answer
--- in terms of substitutions. A substitution is defined as a mapping of terms
to elements from a set of variables. An answer substitution for a query $Q$ is
any substitution $\sigma$ such that
\( \{ x \mid \sigma(x) \not \equiv x \} \subseteq \VARS{Q} \)
holds.

The standard declarative semantics associated with any logic program $P$ is
based on the logical consequence. Given a query $Q$ it declares that an answer
substitution $\sigma$ for $Q$ is a correct answer substitution for
\( P \cup \{ Q \} \)
if $Q\sigma$ is a logical consequence of the program $P$. Here $Q\sigma$
denotes the application of $\sigma$ to $Q$. The declarative semantics provides
a precise definition of the meaning of logic programs, which is independent of
any procedural considerations. It can be served as a specification for
computational mechanism.

On the other side, there are many procedural mechanism for evaluating queries
(bottom-up, top-down, mixed, parallel, \ldots). Each of them implies an
operational semantics of logic programs. A computation mechanism should be
described to produce answers in accordance with the declarative semantics. In
particular, we demand that some completeness results can be proved under some
additional assumptions to a class of queries and programs. When we deal with
substitutions, such (computed) answers are called computed answer
substitutions.

The majority of computation methods in logic programming is based on the notion
of resolvent. Given a query $Q$ and a clause $C$, a resolvent $Q'$ of $Q$ and
$C$ is obtained by performing the following two steps: (i) compute a most
general unifier $\sigma$ of a selected atom in $Q$ and the head of a clause
$C'$, where $C'$ is a renamed clause $C$ such that it has no variables in
common with the selected atom, and (ii) if (i) succeeds, then replace in $Q$
the selected atom by the body of $C'$ and apply the substitution $\sigma$. We
write \RULEA{Q}{}{Q'}. The renaming condition for the clause $C$ is called
standardization apart. The idea is that we do not wish to make the result of
the unification dependent on the choice of variables. Nevertheless, the
semantics of resolvents depends on renaming. We now show that this dependency
implies relevant problems in order to describe operational semantics:
\begin{enumerate}
	\item[(a)]
			SLD-derivation can be defined as the transitive closure of the
			relation \RULEA{}{}{} on queries. Consider a program $P$ and a query 
			$Q$.  By a (partial) SLD-derivation of
			\( P \cup \{ Q \} \) 
			we mean a (possible infinite) sequence 
			\( Q_0, Q_1, Q_2, \ldots \) 
			of queries, where \
			\( Q_0 \equiv Q \), 
			together with a sequence 
			\( C_0, C_1, C_2, \ldots \) 
			of clauses from $P$ and a sequence 
			\( \sigma_0, \sigma_1, \sigma_2, \ldots \) 
			of substitutions such that $Q_{i+1}$ is a resolvent of $Q_i$ and $C_i$
			with {\em mgu\/} $\sigma_i$. The clauses
			\( C_0, C_1, \ldots \) 
			are called input clauses in this derivation. If this derivation is
			finite and the last query $Q_n$ is empty, then a computed answer 
			substitution is obtained as a restriction of the composition
			\( \sigma_0 \ldots \sigma_{n-1} \) 
			to the variables of $Q$.
			
			Because of presence of variables we have to be careful. To ensure
			completeness we need stronger conditions to the process of renaming of
			input clauses in computing resolvents.  Suppose that $C'_i$ denotes a
			clause which is obtained by renaming of $C_i$ in order to compute the
			resolvent $Q_{i+1}$ in this derivation.  For instance, derivations as
			defined in \cite{Apt:logic} require that $C'_{i+1}$ has no variables 
			in common with $Q_0$, $C'_1$, \ldots, $C'_i$.  Moreover all 
			substitutions used in derivations have to be idempotent. We refer to 
			\cite{KN:revisited} for another conditions on ''variable 
			disjointness''. Moreover, in \cite{KN:revisited}, crucial statements 
			which are used in the proof of completeness of SLD-resolution were 
			corrected. As a consequence, we lost ''transitive'' properties of 
			SLD-derivation. In particular, we cannot directly to connect two 
			(partial) derivations to obtain a new one in general, since a 
			required variable-separated condition must not hold.
	\item[(b)]
			There are similar (and more relevant) problems when we deal with
			concurrent models of logic languages. In \cite{Palamidesi:idemp} an
			operation on idempotent substitutions, called parallel composition,
			has been introduced. This operation allows us to combine results from
			parallel execution. The following example shows that this operation
			cannot preserve the completeness in general. Consider for instance
			queries
			\( Q_1 \equiv p(x) \),
			\( Q_2 \equiv q(y) \)
			and a program
			\( \{ \CLAUSE{p(f(z))}{}, \CLAUSE{q(g(z))}{} \} \). 
			If we run $Q_1$ and $Q_2$ in parallel and independently, we could get 
			answers
			\( \sigma_1 = \{ x \leftarrow f(z) \} \) 
			for $Q_1$ and
			\( \sigma_2 = \{ y \leftarrow g(z) \} \) 
			for $Q_2$. 
			By the parallel composition of $\sigma_1$ and $\sigma_2$ we obtain an
			answer
			\( \sigma = \{ x \leftarrow f(z), y \leftarrow g(z) \} \) 
			for 
			\( Q \equiv Q_1 \wedge Q_2 \). 
			Of course an expected answer for $Q$ is a substitution of the form
			\( \{ x \leftarrow f(z_1), y \leftarrow g(z_2) \} \),
			where 
			\( z_1 \not \equiv z_2 \).
			Therefore we need an other, adequate operation on substitutions to
			ensure the completeness of parallel execution models of logic
			programs.
\end{enumerate}

An approach to solve these difficulties is to replace the notion of
substitution by the one of equation set. Namely, the unification algorithm can
be viewed as an equation solving process. Consider the following two atoms
\( A \equiv p(s_1, \ldots, s_n) \) 
and
\( A' \equiv p(t_1, \ldots, t_n) \).
An equation set
\( \{ s_1 = t_1, \ldots, s_n = t_n \} \)
is transformed step by step until a solved form 
\( \{ x_1 = r_1, \ldots, x_k = r_k \} \)
is reached from which a most general unifier of $A$ and $A'$ can be obtained. 
Namely, the substitution
\( \{ \BINDA{x_1}{r_1}, \ldots, \BINDA{x_k}{r_k} \} \)
is an {\em mgu\/} of $A$ and $A'$. A nice theory for equation solving was
developed in \cite{LMM:unif}.

Our approach is based on the equation solving. By allowing existentially
quantified variables in queries we eliminate problems when we deal with new
variables from input clauses.  Namely, suppose that
\( A \equiv p(s_1, \ldots, s_n) \) 
is a selected atom in a query $Q$ and $C$ is a clause. Let
\( C' \equiv \CLAUSE{p(t_1, \ldots, t_n)}{B_1 \wedge \ldots \wedge B_q} \)
be a clause obtained from $C$ by renaming of variables such that it has no
variables in common with $A$. Suppose
\( \VARS{C'} = \{ z_1, \ldots, z_m \} \).
Then we replace $A$ in $Q$ by a formula (called an
atomic reduction of $A$ by $C$) of the form
\[ (\exists z_1) \ldots (\exists z_m) (s_1 = t_1 \wedge \ldots \wedge s_n = t_n
\wedge B_1 \wedge \ldots \wedge B_q) \] 
and then a new query $Q'$ called a resolvent of $Q$ and $C$ is obtained by a
step-by-step transformation according some rewriting rule. The rewriting
process can be viewed as a generalization of the equation solving for formulas
containing atoms, equations, conjunctions and existentially quantified
variables. Notice that the semantics of the atomic reduction is independent on
the choice of names variables in the input clause.

In our formalism we shall deal with a generalization of the standard notion of
query. Queries will be formulas constructed from atoms and equations using the
conjunction and the existential quantifier. An empty query is a query having no
occurrence of atoms. Thus empty queries are just existentially quantified
systems of equations. The role of answers will be played by existentially
quantified systems of equations. We generalize the standard SLD-resolution for
such queries and establish the soundness (Theorem~\ref{cl:sound}) and the
completeness (Theorem~\ref{cl:compl}) of this computation mechanism.  Since we
allow existentially quantified variables in queries, the operational semantics
will be independent on the choice of names of variables in input clauses.
Moreover, the parallel composition of answers can be defined directly through a
(consistent) conjunction of answers.

\section{Notation and Definitions}\label{sc:basic}

In this section we recall some basic definitions. We refer to
\cite{Apt:logic,Lloyd:logic,Shoenfield:logic} for a more detailed presentation
of our topics.

\paragraph{Syntax}
The {\em alphabet\/} $L$ for a first order language consists of {\em logical
symbols\/} (a denumerable set of variables, punctuation symbols, connectives
and quantifiers) and two disjoint classes of nonlogical symbols: (i) a set
\FUNC{\LB{}} of function symbols (including constants) and (ii) a set
\PRED{\LB{}} of predicate symbols. Throughout this paper we assume that the set
of function symbols contains at least one constant. Moreover we always suppose
that the equality symbol $=$ and propositional constants \TRUE\ and \FALSE\ are
contained in all alphabets we shall use. We shall write 
\( e_1 \equiv e_2 \)
to denote the syntactical identity of two strings $e_1$ and $e_2$ of symbols.
We denote by \VAR\ the set of all variables.

We use $u$, $v$, $x$, $y$ and $z$, as syntactical variables which
vary through variables; $f$, $g$ and $h$ as syntactical variables which vary
through function symbols; $p$ and $q$ as syntactical variables which vary
through predicate symbols excluding $=$; and $a$, $b$, $c$ and $d$ as
syntactical variables which vary through constants.

The {\em first order language} consists of two classes of strings of symbols
over a given alphabet $L$: (i) a set of terms, denoted \TERM{\LB{}}, and (ii) a
set of all well-formed formulas. We use $r$, $s$ and $t$, as syntactical
variables which vary through terms; $F$, $G$ and $H$ as syntactical variables
which vary through formulas.  An {\em equation\/} is a formula of the form
$s=t$; and an {\em atom\/} is a formula of the form
\( p(s_1, \ldots, s_n) \).
We use $A$, $B$ and $C$, as syntactical variables which
vary through atoms. A formula is called {\em positive\/} if it is constructed
from propositional constants \TRUE\ and \FALSE, and from equations and atoms
using the conjunction $\wedge$, the disjunction $\vee$ and quantifiers. By an
{\em equational\/} formula we mean any formula having no occurrence of atoms.

Consider a term $s$. Then \VARS{s} denotes the set of variables appearing in
$s$. If \VARS{s} is empty, then the term $s$ is called {\em ground\/}.
Similarly, \VARS{F} denotes the set of free variables of a formula $F$. $F$ is
said to be {\em closed\/} if
\( \VARS{F} = \emptyset \).
Let
\( x_1, \ldots, x_n \)
be all distinct variables occurring freely in  a formula $F$ in this order. We
write
\( (\forall) F \)
or
\( (\exists) F \)
for
\( (\forall x_1) \ldots (\forall x_n) F \)
or
\( (\exists x_1) \ldots (\exists x_n) F \),
respectively. We call
\( (\forall) F \)
or
\( (\exists) F \)
the {\em universal closure\/} or the {\em existential closure\/} of $F$.

In order to avoid the awkward expression {\em the first order language over an
alphabet \LA{}\/} we will simple say {\em the first order language \LA{}\/} (or
{\em the language \LA{}\/} for short).

\paragraph{}
In Section~\ref{sc:semadecl} we shall use the notion of application of finite
substitutions to formulas. The definition is based on a weaker form of this
application. We follow here \cite{Shoenfield:logic}:
\begin{enumerate}
	\item [(a)] 
			\( t_{x_1, \ldots, x_n}[s_1, \ldots, s_n] \) 
			denotes a term obtained from $t$ by {\em simultaneously\/} replacing 
			of each occurrence
			\( x_1, \ldots, x_n \) 
			in $t$ by 
			\( s_1, \ldots, s_n \),
			respectively.
	\item [(b)]
			\( F_{x_1, \ldots, x_n}[s_1, \ldots, s_n] \) 
			denotes a formula obtained from $F$ by {\em simultaneously\/} 
			replacing of each {\em free\/} 
			occurrence 
			\( x_1, \ldots, x_n \) 
			in $F$ by 
			\( s_1, \ldots, s_n \),
			respectively.
\end{enumerate}
Whenever
\( t_{x_1, \ldots, x_n}[s_1, \ldots, s_n] \) 
or
\( F_{x_1, \ldots, x_n}[s_1, \ldots, s_n] \) 
appears,
\( x_1, \dots , x_n\) 
are restricted to represent distinct variables. Moreover, in (b) we always
suppose that each term $s_i$ is {\em substitutible\/} for $x_i$ in $F$
i.e. for each variable $y$ occurring in $s_i$, no part of $F$ of the form
\( (\exists y) G \) (or \( (\forall y) G \) ) 
contains an occurrence of $x_i$ which is free in $F$. We shall omit the
subscripts
\( x_1, \ldots, x_n \)
when they occur freely in $F$ in this order and
\( \VARS{F} = \{ x_1, \ldots, x_n \} \).

We say that $F'$ is a {\em variant\/} of $F$, if $F'$ can be obtained
from $F$ by a sequence of replacement of the following type: replace a
part
\( (\exists x) G \) 
or 
\( (\forall x) G \)
by 
\( (\exists y) G_x[y] \) 
or by
\( (\forall x) G_x[y] \),
respectively, where $y$ is a variable not free in $G$.

\paragraph{}
By {\em free equality axioms\/} for a language \LA{} 
(see \cite{Apt:logic,Lloyd:logic}), 
we mean the theory \FEA{\LB{}} consisting of the following formulas:
\begin{enumerate}
	\item[(a)]
			\( f(x_1, \ldots, x_n) = f(y_1, \ldots, y_n) \leftrightarrow 
			x_1 = y_1 \wedge \ldots \wedge x_n = y_n \)
			for each $n$-ary function symbol $f$,
	\item[(b)]
			\( f(x_1, \ldots, x_n) = g(y_1, \ldots, y_m) \leftrightarrow \FALSE \)
			for each $n$-ary function symbol $f$ and $m$-ary function symbol 
			$g$ such that 
			\( f \not \equiv g \),
	\item[(c)]
			\( x = t \leftrightarrow \FALSE \)
			for each variable $x$ and term $t$ such that 
			\( x \not \equiv t \) 
			and $x$ occurs in $t$.
\end{enumerate}
Since we identify constants with $0$-ary function symbols, then (b) includes 
\( a \not = b \) 
for pairs of distinct constants as a special case.

\paragraph{Semantics}
A {\em pre-interpretation\/} $J$ for a language \LA{}\ consists of (i) a
non-empty universe \UU{L}{J}, called a {\em domain\/} of $J$, and (ii) a fixed
interpretation of all function symbols. The equality $=$ is interpreted as the
identity on \UU{L}{J}.  An {\em interpretation\/} $I$ for \LA{}\ is based on
$J$ (or just {\em J-interpretation} for short) if it is obtained from $J$ by
selecting some interpretation of predicate symbols.

Consider a pre-interpretation $J$. A variable assignment
\( h : \VAR \rightarrow \UU{L}{J} \) 
will be called a {\em valuation\/} over $J$ (or a {\em J-valuation\/} for
short).  The set of all {\em J\/}-valuations is denoted by \VV{L}{J}. Obviously
each {\em J\/}-valuation $h$ has a unique homomorphic extension $h'$ from
\TERM{} into \UU{L}{J}. We shall write $h(s)$, where $s$ is a term, instead of
$h'(s)$. We call $h(s)$ a {\em J-instance\/} of $s$. Consider an atom 
\( A \equiv p(s_1, \ldots, s_n) \)
and a {\em J\/}-valuation $h$. The generalized atom 
\( h(A) \equiv p(h(s_1), \ldots, h(s_n)) \).
will be called a {\em J-instance\/} of $A$. The set of all {\em J\/}-instances
of atoms, called {\em J-base\/}, is denoted by \BB{L}{J}. We shall identify
interpretations based on $J$ with subsets of \BB{L}{J}.

\begin{example}\label{ex:herbrand}
	The {\em Herbrand\/} pre-interpretation \HA\ for $L$ is defined as follows:
	\begin{enumerate}
		\item[(a)]
				Its domain is the set \UU{L}{\HB} of all ground terms of $L$;
				called the {\em Herbrand universe\/}.
		\item[(b)]
				Each constant in $L$ is assigned to itself.
		\item[(c)]
				If $f$ is an {\em n\/}-ary function symbol in $L$ then it is
				assigned to the mapping from $(U_L^{\HB})^n$ to \UU{L}{\HB} defined
				by assigning the ground term
				\( f(s_1, \ldots, s_n) \)
				to the sequence
				\( s_1, \ldots, s_n \)
				of ground terms.
	\end{enumerate}
	By a {\em Herbrand\/} interpretation for $L$ we mean any interpretation
	based on \HA. As remarked above we shall identify Herbrand interpretations
	with subsets of the set \BB{L}{\HB} called the {\em Herbrand base\/}.
\end{example}

Given a formula $F$ we define its {\em truth\/} in a $J$-valuation $h$ and an
interpretation $I$ based on $J$, written as \TRUTH{I}{h}{F}, in the obvious
way.  In particular, \TRUTH{I}{h}{s = t} iff
\( h(s) \equiv h(t) \).
We call
\( h \in \VV{\LB{}}{J} \)
a solution of $F$ in $I$, if \TRUTH{I}{h}{F}. The set of solutions of $F$ in
$I$ is denoted by \ANS{F}{I}{J}. So:
\[	\ANS{F}{I}{J} = \{ h \in \VV{\LB{}}{J} \mid \TRUTH{I}{h}{F} \} \]
We say that a formula $F$ is {\em true\/} in $I$, written as \TRUTH{I}{}{F},
when for all valuations
\( h \in \VV{\LB{}}{J} \), 
\TRUTH{I}{h}{F}. We say that a formula is {\em valid\/} when it is true in any
interpretation. 

Consider a theory $T$ over \LA{}. An interpretation $I$ is a {\em model\/} for
$T$ if \TRUTH{I}{}{F} for any $F$ from $T$. A formula $F$ (over \LA{}) is  a
{\em logical consequence\/} of $T$ if any model $I$ for $T$ is a model for $F$,
as well. We write \TRUTH{T}{}{F} in \LA{} (or \TRUTH{T}{}{F} for short).

Consider an equation formula $F$. Note that the semantics of $F$ depends only
on an interpretation of function symbols. Consequently, given a
pre-interpretation $J$, the truth of $F$ in a $J$-valuation $h$ is
well-defined. We shall write \TRUTH{J}{h}{F}. We call
\( h \in \VV{\LB{}}{J} \)
a solution of $F$ in $J$ (or {\em $J$-solution\/} of $F$ for short), if
\TRUTH{J}{h}{F}. The set of $J$-solutions of $F$ is denoted by \SOLN{J}{F}. So:
\[ \SOLN{J}{F} = \{ h \in \VV{\LB{}}{J} \mid \TRUTH{J}{h}{F} \} \]
$F$ is said to be {\em true\/} in $J$, written as \TRUTH{J}{}{F}, when for all
valuations
\( h \in \VV{L}{J} \), 
is \TRUTH{J}{h}{F}. We say that $J$ is a model for \FEA{L}, when any axiom of
\FEA{L} is true in $J$.

The following theorems will be used in the sequel (see
\cite{Shoenfield:logic}).

\begin{theorem}[Variant Theorem]\label{cl:variant}
	If $F'$ is a variant of $F$, then 
	\( F' \leftrightarrow F \)
	is valid.
\end{theorem}

\begin{theorem}[Theorem on Constants]\label{cl:constants}
	Let $T$ be a theory and $F$ a formula. If 
	\( a_1, \dots, a_n \) 
	are distinct constants not occurring in $T$ and $F$, then 
	\[ \TRUTH{T}{}{F}\ \ {\rm iff}\ \ \TRUTH{T}{}{F[a_1, \ldots, a_n]} \]
\end{theorem}

\paragraph{Logic programs and queries}
A {\em conjunctive query\/} is a formula constructed from propositional
constants \TRUE\ and \FALSE, and from equations and atoms using the conjunction
$\wedge$, the negation $\neg$ and the existential quantifier $\exists$. A {\em
clause\/} $C$ is a formula of the form \CLAUSE{A}{Q}, where $A$ is an atom,
called a {\em head\/} of $C$ and $Q$ is a query, called a {\em body\/} of $C$.
If the body of $C$ is a positive query, it is called a {\em definite\/} clause.
By a program $P$ we mean any finite set of clauses. If a program contains only
definite clauses, then it is called a {\em definite\/} program (or a {\em
positive\/} program).  We denote by $P^*$, where $P$ is a program over a
language \LA{}, the theory over \LA{P} obtained by adding free equality axioms
for the language \LA{} to the program $P$.

In this paper we deal only with positive queries and positive programs. From
now by a query or a program we shall always mean a positive query or a positive
program, respectively. We denote by \CARD{Q} the number of atoms occurring
in $Q$. If
\( \CARD{Q} = 0 \),
then $Q$ is said to be {\em empty\/}. A clause will called a {\em fact\/} or
{\em rule\/} it its body is an empty or a nonempty query, respectively.

Let $J$ be a pre-interpretation for \LA{} and $P$ a program. By $J$-model for
$P$ we mean any model for $P$ based on $J$.  Clearly \BB{L}{J} is a $J$-model
for $P$. Moreover, the set of $J$-models for $P$ is closed under (set)
intersection. The intersection \MM{P}{J} of all $J$-models for $P$ is thus a
model, as well. The least $J$-model \MM{P}{J} of $P$ can be obtained by
iteration of an immediate consequence operator \TOA{P}{J} defining on the set
of $J$-interpretations. The {\em immediate consequence operator\/} \TOA{P}{J}
maps $J$-interpretations to $J$-interpretations as follows:
\begin{quote}
	\( p(d_1, \ldots, d_n) \in \TOBB{P}{J}{I} \)
	iff for some
	\( h \in \VV{\LB{}}{J} \)
	and a clause \CLAUSE{A}{Q} of $P$ we have
	\( p(d_1, \ldots, d_n) \equiv h(A) \)
	and \TRUTH{I}{h}{Q}.
\end{quote}
We denote by
\( lfp(\TOA{P}{J}) \)
or \TOCA{P}{J}{\alpha} the last fixpoint of \TOA{P}{J} or the upward power for
an ordinal $\alpha$, respectively. The following well-known theorem holds.

\begin{theorem}\label{cl:characterization}
	Let $P$ be a program over \LA{} and $J$ a pre-interpretation for \LA{}. 
	Then:
	\begin{enumerate}
		\item[(a)]
				\TOA{P}{J} is a continuous operator
		\item[(b)]
				$I$ is a model for $P$ iff 
				\( \TOBB{P}{J}{I} \subseteq I \)
		\item[(c)]
				\( \MM{P}{J} = \TOCB{P}{J}{\omega} = lfp(\TOA{P}{J}) \)
	\end{enumerate}
\end{theorem}

{\bf Proof:}
Standard (see \cite{Apt:logic,Lloyd:logic}).
\END

\medskip

A query $Q$ is said to be in {\em solved form\/} if 
\begin{enumerate}
	\item[(a)]
			$Q$ is a propositional constant \TRUE\ or \FALSE\ or
	\item[(b)]
			$Q$ is of the form
			\[ (\exists z_1) \ldots (\exists z_k) (x_1 = s_1 \wedge \ldots \wedge 
				x_n = s_n \wedge A_1 \wedge \ldots \wedge A_m) \],
			where (i) $x_i$'s and $z_j$'s are distinct variables, (ii) $x_i$'s
			occur nor in the right hand side of any equation nor in any atom,
			(iii) each $z_j$ has an occurrence in the conjunction and (iv) 
			\( z_j \not \equiv s_i \) 
			for any $i$ and $j$.
\end{enumerate}
The variables 
\( x_1, \ldots, x_n \) 
in (b) are said to be {\em eliminable} and the set 
\( \{ x_1, \ldots, x_n \} \) 
is denoted by \ELIM{Q}. The remaining free variables in $Q$ are called {\em
parameters} and the set of parameters is denoted by \PARAM{Q}. So 
\( \VARS{Q} = \ELIM{Q} \cup \PARAM{Q} \). 
The set 
\( \{ z_1, \ldots, z_k \} \) 
of (existentially) bound variables in $Q$ is denoted by \BOUND{Q}. We put 
\( \ELIM{\TRUE} = \PARAM{\TRUE} = \BOUND{\TRUE} = \emptyset \).

We say that a query $Q$ is {\em consistent\/} if there is an interpretation $I$
(based on a pre-interpretation $J$), which is a model for \FEA{\LB{}}, such
that \TRUTH{I}{h}{Q} for some
\( h \in \VV{\LB{}}{J} \),
or equivalently if \NOTTRUTH{\FEA{\LB{}}}{}{\neg Q}. Note that the consistency
of queries in solved form can be checked directly: if $Q$ is a query in solved
form, then $Q$ is consistent iff
\( Q \not \equiv \FALSE \).
In Section~\ref{sc:opersema} we shall describe an algorithm, called Solved Form
Algorithm, which transforms every query into a solved form. As we see later 
(Theorem~\ref{cl:Robinson:1}), the algorithm preserves consistency. Hence the
consistency of a query can be reached directly by the the form of its
(computed) solved form.

We write that \LA{} is \LA{P}, if \LA{} is a first order language associated
with a program $P$. Then we denote \UU{\LB{}}{J}, \BB{\LB{}}{J}, \VV{\LB{}}{J},
\ldots by \UU{P}{J}, \BB{P}{J}, \VV{P}{J}, \ldots, respectively. Since we
consider only alphabets having at least one constants, as a consequence we have
that the Herbrand universe \UU{P}{\HB} and the Herbrand base \BB{P}{\HB} are
nonempty sets. In the sequel, we shall deal with extensions of a given first
order language. We say that the first order language \LA{2} is an {\em
extension\/} of the the first order language \LA{1} if every nonlogical symbol
of \LA{1} is a nonlogical symbol of \LA{2}. The following claim will be used
later.

\begin{lemma}[\cite{Shoenfield:logic}]\label{cl:extension}
	Consider a theory $T$ over \LA{1}. If \LA{2} is obtained from \LA{1} by
	adding some constants, then
	\[ \TRUTH{T}{}{F}\ {\rm in}\ \LB{1}\ \ {\rm iff}\ \ 
		\TRUTH{T}{}{F}\ {\rm in}\ \LB{2} \]
	for any formula $F$ over \LA{1}.
\end{lemma}

\noindent
We restrict our attention only to extensions which are obtained by adding
denumerable number of new constants. We say that the first order language
\LA{c} is a {\em canonical language\/} for a program $P$ if \LA{c} is obtained
by adding denumerable number of new constants to \LA{P}.

Consider a clause 
\( C \equiv \CLAUSE{A}{Q} \).
We say that a clause $C'$ is obtained from $C$ by {\em renaming\/} of variables
in $C$ if
\( C' \equiv C_{x_1, \ldots, x_n}[y_1, \ldots, y_n] \),
where
\( \{ x_1, \ldots, x_n \} = \VARS{C} \)
and $y_i$'s are pairwise distinct variables. Then
\( \TOBB{C'}{J}{I} = \TOBB{C}{J}{I} \)
for any interpretation $I$ based on $J$. Consequently the operator \TOA{P}{J}
is independent on the choice of names of variables in $P$.

\section{Declarative Semantics}\label{sc:semadecl}

In this section we provide a link between the explicit and the implicit
representation of answer. The role of the explicit representation of answer
will be played by finite substitutions and the role of the implicit one by
existentially quantified systems of equations. We follow here
\cite{Komara:unif}.

\paragraph{Finite substitutions}
By a {\em finite substitution\/} we mean any mapping of terms to variables from
a finite set $X$ of variables. Let
\( X = \{ x_1, \ldots, x_n \} \).
We shall use the standard set-theoretic notation 
\( \sigma = \{ \BINDA{x_1}{s_1}, \ldots, \BINDA{x_n}{s_n} \} \) 
to denote $\sigma$, where
\( s_i \equiv \sigma(x_i) \).
We say that $\sigma$ is {\em over\/} $X$. We denote by \DOM{\sigma} or
\RANGE{\sigma} the set $X$ or the set of variables occurring in terms
\( s_1, \ldots, s_n \),
respectively. The pair \BINDA{x}{s} is called a {\em binding}. In the sequel by
a substitution we always mean a finite substitution.

Let $\sigma$ and $\theta$ are substitutions with disjoint domains. By
\( \sigma \cup \theta \)
we mean a substitution over
\( \DOM{\sigma} \cup \DOM{\theta} \)
assigning
\( \sigma(x) \)
or
\( \theta(x) \)
to
\( x \in \DOM{\sigma} \)
or
\( x \in \DOM{\theta} \),
respectively. We call $\sigma$ a {\em permutation\/} if it is one-to-one
mapping from \DOM{\sigma} onto \RANGE{\sigma}. There is unique substitutions
\SUBEMPTYA\ with the empty domains; it will be called the {\em empty\/}
substitution.

A substitution
\( \sigma = \{ \BINDA{x_1}{s_1}, \ldots, \BINDA{x_n}{s_n} \} \)
is {\em applicable\/} to a term $t$ if \DOM{\sigma}
contains all variables occurring in $t$. Then an {\em application\/} of
$\sigma$ to $t$, denoted $t \sigma$,
is defined as the term
\( t_{x_1, \ldots, x_n}[s_1, \ldots, s_n] \).
The term $t \sigma$ is called an {\em instance of\/} $t$. If $\sigma$ is a
permutation, then $t \sigma$ and $t$ are said to be {\em variants\/}.

A substitution
\( \sigma = \{ \BINDA{x_1}{s_n}, \ldots, \BINDA{x_n}{s_n} \} \)
is {\em applicable\/} to a formula $F$ if each $s_i$ is substitutible for $x_i$
in $F$. Then an {\em application\/} of $\sigma$ to $F$, denoted
\( F \sigma \),
is defined as a formula
\( F_{x_1, \ldots, x_n}[s_1, \ldots, s_n] \).

A substitution $\theta$ is {\em applicable\/} to $\sigma$ if 
\( \RANGE{\sigma} \subseteq \DOM{\theta} \).
Then a {\em composition\/} of $\sigma$ and $\theta$ is a substitution over
\DOM{\sigma}, denoted
\( \sigma \theta \),
and it is defined in the obvious way. If $\theta$ is a permutation, then we say
that $\sigma$ and $\sigma \theta$ are {\em variants\/}.

We denote by \RESTR{\sigma}{X} a {\em restriction\/} of $\sigma$ onto $X$. So
\RESTR{\sigma}{X} is a substitution over
\( \DOM{\sigma} \cap X \).
We call $\sigma$ an {\em extension\/} of \RESTR{\sigma}{X} onto \DOM{\sigma}.
By a {\em regular extension\/} of $\sigma$ onto $X$, where
\( \DOM{\sigma} \subseteq X \),
we mean any extension $\theta$ of $\sigma$ onto $X$ such that $\theta$ maps the
set
\( X \setminus \DOM{\sigma} \)
injectively into the set
\( \VAR \setminus \RANGE{\sigma} \)
of variables.

The preorder $\preceq$ and the equivalence $\approx$ on substitutions is based
on the notion of composition:
\begin{enumerate}
	\item[(a)]
			$\theta$ is {\em more general} than $\sigma$, written as 
			\( \sigma \preceq \theta \), 
			if there is $\tau$ such that 
			\( \sigma' = \theta' \tau \), 
			where $\sigma'$ and $\theta'$ are regular extensions of $\sigma$ and 
			$\theta$ respectively over the same domain 
	\item[(b)]
			$\theta$ is {\em equivalent} to $\sigma$, written as 
			\( \theta \approx \sigma \), 
			if 
			\( \theta \preceq \sigma \) 
			and 
			\( \sigma \preceq \theta \).
\end{enumerate}
Note that, if $\sigma$ and $\theta$ are variants, then they are equivalent.
Moreover (see
\cite{Komara:unif}):
\begin{enumerate}
	\item[(a)]
			if 
			\( \sigma \preceq \theta \),
			then
			\( \RESTR{\sigma}{X} \preceq \RESTR{\theta}{X} \)
	\item[(b)]
			if 
			\( \sigma \approx \theta \),
			then
			\( \RESTR{\sigma}{X} \approx \RESTR{\theta}{X} \)
\end{enumerate}
Clearly equivalent substitutions may have different domains.  Nevertheless, to
any substitutions $\sigma$ we can find a minimal set $X$ of variables (under
set inclusion), denoted \KERNEL{\sigma}, having the following property (see
\cite{Komara:unif}): the restriction of $\sigma$ onto $X$ is equivalent to
$\sigma$. Then (\cite{Komara:unif}):
\begin{enumerate}
	\item[(a)]
			$\sigma$ is a regular extension of \RESTR{\sigma}{\KERNEL{\sigma}}
	\item[(b)]
			if 
			\( \sigma \approx \theta \),
			then
			\( \KERNEL{\sigma} = \KERNEL{\theta} \)
\end{enumerate}
We shall use \KERNEL{\ldots} to define the notion of answer substitution.

Let \SUBFALSE\ be an arbitrary object that is not element of \SUBA{L}. Let
\SUBB{L} be the set 
\( \SUBA{L} \cup \{ \SUBFALSE \} \).
We extend the preorder $\preceq$ and the equivalence $\approx$ to
\SUBB{L}\ by requiring \SUBFALSE\ to be the smallest element of \SUBB{L}. We 
denote by \SUBC{L}\ the new quotient set and by $\sigma_{\approx}$ the
equivalence class in which $\sigma$ lies. Then \SUBC{L}\ is a complete lattice
with \SUBEMPTYB\ as the greatest element and with \SUBFALSE\ as the smallest
element (see \cite{Komara:unif}).

In \cite{Komara:unif} the notion of application of substitutions to formulas
was generalized for arbitrary substitutions. The following claim holds.

\begin{proposition}[\cite{Komara:unif}]\label{cl:application}
	The application of substitutions to formulas has the following properties:
	\begin{enumerate}
		\item[(a)]
				If $F$ and $F'$ are variants, then 
				\( F \sigma \)
				and
				\( F' \sigma \)
				are variants, as well.
		\item[(b)]
				If
				\( \sigma \preceq \theta \),
				then
				\( (\forall) F \sigma \rightarrow (\forall) F \theta \)
				is valid.
		\item[(c)]
				If
				\( \sigma \approx \theta \),
				then
				\( (\forall) F \sigma \leftrightarrow (\forall) F \theta \)
				is valid.
		\item[(d)]
				Let
				\( \sigma = \{ \BINDA{x_1}{s_n}, \ldots, \BINDA{x_n}{s_n} \} \)
				be applicable to $F$. Then
				\[ F \sigma = F_{x_1, \ldots, x_n}[s_1, \ldots, s_n]. \]
	\end{enumerate}
\end{proposition}

Consider a program $P$ and a query $Q$. We say that a substitution $\sigma$ is
an {\em answer substitution\/} (or just an {\em answer\/} for short) for $Q$ if
\( \KERNEL{\sigma} \subseteq \VARS{F} \). 
Moreover, if \TRUTH{P}{}{Q\sigma}, then $\sigma$ is called a {\em correct
answer substitution\/} (or {\em correct answer\/} for short) for
\( P \cup \{ Q \} \).

\paragraph{Existentially quantified systems of equations}
In our formalism we shall deal with an implicit representation of answer. The
role of answer will be played existentially quantified systems of equations.
By a (positive) {\em existentially quantified system of equations\/} (or {\em
\EQ{formula\/} for short}) we mean any formula constructed from propositional
constants \TRUE\ and \FALSE, and from and equations using the conjunction
$\wedge$ and the existential quantifier $\exists$. Thus \EQ{formulas} are just
empty queries.

In \cite{Komara:unif} the preorder $\preceq$ and the equivalence $\approx$ on
\EQ{formulas} was introduced:
\begin{enumerate}
	\item[(a)]
			$E'$ is more general than $E$, written as
			\( E \preceq E' \), 
			if \TRUTH{\FEA{L}}{}{E \rightarrow E'}
	\item[(b)]
			$E'$ is equivalent to $E$, written as
			\( E \approx E' \), 
			if \TRUTH{\FEA{L}}{}{E \leftrightarrow E'}
\end{enumerate}
It was shown that the quotient set \EQNB{\LB{}} is a complete lattice, where
the topmost element is the class modulo $\approx$ containing the propositional
constant \TRUE\ and the lowest element represents inconsistent \EQ{formulas},
in particular, it contains the propositional constant \FALSE.

Another definition of a preorder and an equivalence on equation sets was
introduced in \cite{LMM:unif}.  Actually their notions strictly corresponds to
a preorder $\preceq_{\HB}$ and an equivalence $\approx_{\HB}$, where \HA\ is a
Herbrand pre-interpretation, in the following sense. Let $J$ be a
pre-interpretation for a language $L$. Then:
\begin{enumerate}
	\item[(a)] 
			\( E \preceq_J E' \) 
			if \TRUTH{J}{}{E \rightarrow E'} or equivalently if 
			\( \SOLN{J}{E} \subseteq \SOLN{J}{E'} \).
	\item[(b)]
			\( E \approx_J E' \)
			if \TRUTH{J}{}{E \leftrightarrow E'} or equivalently if 
			\( \SOLN{J}{E} = \SOLN{J}{E'} \).
\end{enumerate}
We say that $J$ is a {\em non-trivial\/} pre-interpretation if \UU{L}{J} has at
least two elements. The following proposition holds.

\begin{proposition}[\cite{Komara:unif}]\label{cl:order}
	Let $E$ and $E'$ be \EQ{formulas}. If $J$ is a non-trivial model of
	\FEA{L}, then: 
	\begin{enumerate} 
		\item[(a)]
				\( E \approx_J E' \) iff \( E \approx E' \) 
		\item[(b)]
				\( E \preceq_J E' \) iff \( E \preceq E' \) 
	\end{enumerate} 
\end{proposition}

The following property of equivalent consistent \EQ{formulas} $E$ and $E'$ both
in solved from was proved in \cite{Komara:unif}:
\begin{enumerate}
	\item[(a)]
			\( E \leftrightarrow E' \) 
			is valid formula
	\item[(b)]
			\( \CARD{\ELIM{E'}} = \CARD{\ELIM{E}} \),
			\( \CARD{\PARAM{E'}} = \CARD{\PARAM{E}} \) 
			and 
			\( \VARS{E'} = \VARS{E} \)
\end{enumerate}
In Section~\ref{sc:opersema} we describe an algorithm, called Solved Form
Algorithm, which transforms any \EQ{formula} into an equivalent one in solved
form. Thus for a given consistent \EQ{formula} $E$ there is a unique set of
variables, denoted by \KERNEL{E}, which is the set of free variables of its
arbitrary solved form. We use this set to define the notion of answer for
\EQ{formulas}.  We say that a consistent \EQ{formula} $E$ is an {\em answer
\EQ{formula}\/} (or just an {\em answer\/} for short) for a query $Q$ if
\( \KERNEL{E} \subseteq \VARS{Q} \).
Moreover, if \TRUTH{P^*}{}{E \rightarrow Q}, then $E$ is called {\em correct
answer \EQ{formula}\/} (or {\em correct answer\/} for short) for
\( P \cup \{ Q \} \).

\paragraph{}
In \cite{Komara:unif} it was proved that the lattice of finite substitutions
\SUBC{\LB{}} is isomorphic to the lattice of \EQ{formulas} \EQNB{\LB{}}. The
crux of this relationship is the mapping between an \EQ{formula}
\[
	E \equiv (\exists z_1) \ldots (\exists z_m) (x_1 = s_1 \wedge \ldots x_n = 
	s_n), 
\]
in solved form and a substitution
\[
	\sigma \equiv \{ \BINDA{x_1}{s_1}, \ldots, \BINDA{x_n}{s_n},
	\BINDA{y_1}{y_1}, \ldots, \BINDA{y_k}{y_k} \}, 
\]
where
\( \{ y_1, \ldots, y_k \} \) 
are parameters of $E$. This mapping has the following property.  If a
substitution $\sigma$ corresponds to an \EQ{formula} $E$ in the isomorphic
mapping between the both lattices, then
\begin{equation}\label{eq:answer:1}
	\KERNEL{\sigma} = \KERNEL{E}
\end{equation}
\begin{equation}\label{eq:answer:2}
	\TRUTH{\FEA{\LB{}}}{}{(\forall) F \sigma \leftrightarrow 
	(\forall) (E \rightarrow F)}
\end{equation}
In particular, $\sigma$ is an answer substitution for $F$ iff $E$ is an answer
\EQ{formula} for $F$. The following theorem establishes a closer link between
the both notion of correct answer.

\begin{theorem}\label{cl:answer:link}
	Consider a program $P$ and a query $Q$. Suppose that a substitution $\sigma$
	corresponds to an \EQ{formula} $E$. Then
	\[ \TRUTH{P}{}{Q \sigma}\ \ {\rm iff}\ \ \TRUTH{P^*}{}{E \rightarrow Q} \]
	In particular, $\sigma$ is a correct answer substitution for
	\( P \cup \{ Q \} \)
	if only if $E$ is an correct answer \EQ{formula} for
	\( P \cup \{ Q \} \).
\end{theorem}

The proof of the claim is based on the equivalence
\( (a) \Leftrightarrow (b) \)
in the following proposition which proof is contained in the Appendix.

\begin{proposition}\label{cl:F:canonical}
	Consider a program $P$ and a positive formula $F$. Let \LA{c} be a 
	canonical language for $P$ and \HA\ a Herbrand pre-interpretation for 
	\LA{c}. Then the following are equivalent:
	\begin{enumerate}
		\item[(a)] \TRUTH{P}{}{F} in \LA{P}
		\item[(b)] \TRUTH{P^*}{}{F} in \LA{P}
		\item[(c)] \TRUTH{\MM{P}{\HB}}{}{F}
		\item[(d)] \TRUTH{\TOCB{P}{\HB}{n}}{}{F} for some $n$
	\end{enumerate}
\end{proposition}

{\bf Proof of Theorem~\ref{cl:answer:link}:}
Straightforward by noting that
\TRUTH{P}{}{Q \sigma}
iff
\TRUTH{P^*}{}{Q \sigma}
by Proposition~\ref{cl:F:canonical} and
\TRUTH{P^*}{}{Q \sigma}
iff
\TRUTH{P^*}{}{E \rightarrow Q}
by~(\ref{eq:answer:2}).
\END

\medskip

In the sequel we shall need the following theorem.

\begin{theorem}\label{cl:E:canonical}
	Consider a program $P$, a query $Q$ and an \EQ{formula} $E$. Let \LA{c} be 
	a canonical language for $P$ and \HA\ a Herbrand pre-interpretation for 
	\LA{c}. Then the following are equivalent:
	\begin{enumerate}
		\item[(a)] \TRUTH{P^*}{}{E \rightarrow Q} in \LA{P}
		\item[(b)] \TRUTH{\MM{P}{\HB}}{}{E \rightarrow Q}
		\item[(c)] \TRUTH{\TOCB{P}{\HB}{n}}{}{E \rightarrow Q} for some $n$
	\end{enumerate}
\end{theorem}

{\bf Proof:}
Let $\sigma$ be a substitution corresponding to $E$. Then by
Theorem~\ref{cl:answer:link} we have \TRUTH{P}{}{Q\sigma} iff 
\TRUTH{P^*}{}{E \rightarrow Q}. (\ref{eq:answer:2}) implies that
\TRUTH{\MM{P}{\HB}}{}{Q\sigma} iff \TRUTH{\MM{P}{\HB}}{}{E \rightarrow Q} and
\TRUTH{\TOCB{P}{\HB}{n}}{}{Q\sigma} iff \TRUTH{\TOCB{P}{\HB}{n}}{}{E 
\rightarrow Q}. The claim now follows from Proposition~\ref{cl:F:canonical}.
\END

\section{Operational Semantics}\label{sc:opersema}

In this section we propose a computation process, a generalization of the
standard SLD-resolution, for queries containing equations and existentially
quantified variables. Queries are computed through a combination of two
mechanisms --- reduction and unification. First we select one or more atoms in
a query $Q$ and then each selected atom is replaced simultaneously in $Q$ by
its atomic reduction. This will be the reduction step.  Then we simplify the
result of the previous step according some rewriting rule possible into a
solved form. So obtained query $Q'$ will be called a resolvent of $Q$.

\paragraph{}
Now we precisely describe one computation step of our procedural mechanism.
Each step will be divided into two elementary actions --- computing a reduction
and unification. The first action is based on the notion of atomic reduction.
Consider an atom
\( A \equiv p(s_1, \ldots, s_n) \) 
and a clause $C$. Let
\( C' \equiv \CLAUSE{p(t_1, \ldots, t_n)}{Q} \) 
be a variant of $C$ having no free variables in
common with $A$. Suppose
\( \{ z_1, \ldots, z_m \} = \VARS{C'} \).
Then the following query
\[	(\exists z_1) \ldots (\exists z_m) (s_1 = t_1 \wedge \ldots \wedge s_n = t_n
	\wedge Q) \]
is called an {\em atomic reduction\/} of $A$ by $C$. The variable-separated
condition is called {\em standardization apart\/}. Notice that the semantics of
the atomic reduction is independent on the choice of new variables. Really, if
queries $Q'$ and $Q''$ are atomic reductions of $A$ by $C$ obtained by possible
different choices of new variables, then they are variants and hence
semantically equivalent i.e.
\( Q' \leftrightarrow Q'' \)
is valid.

Now consider a program $P$ and a query $Q$. Let
\( A_1, \ldots, A_k \)
be selected atoms in $Q$ and let
\( \OLB{C}{} = C_1, \ldots, C_k \)
be a sequence of clauses from $P$ such that $A_i$ and the head of $C_i$ have
the same predicate symbol for each
\( i = 1, \ldots, k \).
Then an {\em reduction\/} of $Q$ by \OLA{C}{} is obtained by simultaneously
replacing of each $A_i$ in $Q$ by its atomic reduction by the clause $C_i$.
From the remark above we have that the semantics of the reduction is
independent on the choice of new variables. More precisely, if queries $Q'$ and
$Q''$ are reductions of $Q$ obtained by possible different choices of new
variables, then they are variants and hence semantically equivalent.

As we marked above the second action is based on unification or more precisely
on query solving. We present an algorithm based upon Solved Form Algorithm
investigated in \cite{LMM:unif} which transforms any query into a solved form.

\paragraph{Solved Form Algorithm}
For a given query $Q$ non-deterministically apply the following elementary
steps (1) - (12). We write
\( (\exists \OLB{y}{}) \)
instead of
\( (\exists y_1) \ldots (\exists y_k) \),
if \OLA{y}{} is the sequence
\( y_1, \ldots, y_k \).

The first group of elementary actions is determined by the form of a selected
equation in $Q$.
\begin{enumerate}
	\item[(1)]
			\( f(s_1, \ldots, s_n) = f(t_1, \ldots, t_n) \)
			replace by
			\( s_1 = t_1 \wedge \ldots \wedge s_n = t_n \)
	\item[(2)]
			\( f(s_1, \ldots, s_n) = g(t_1, \ldots, t_m) \)
			replace by \FALSE, if $f$ and $g$ are distinct symbols
	\item[(3)]
			\( x = t \)
			replace by \FALSE, if $x$ and $t$ are distinct terms such that $x$ 
			occurs in $t$
	\item[(4)]
			\( x = x \)
			replace by \TRUE
\end{enumerate}
The following two actions eliminate a variable $x$ if the selected equation is
of the form $x=t$, where
\( x \not \equiv t \)
and $x$ does not occur in $t$. We suppose that $x=t$ is ''surrounded'' only by
atoms and equations i.e. there is a subquery
\( (\exists \OLB{y}{}) (Q' \wedge x = t \wedge Q'') \)
of the query $Q$, where $Q'$ and $Q''$ are conjunctions only of equations and
atoms. The third action redirects $t=x$ according the form of $t$.
\begin{enumerate}
	\item[(5)]
			\( (\exists \overline{y}) (Q' \wedge x = t \wedge Q'') \)
			replace by
			\( (\exists \overline{y}) (Q'_x[t] \wedge x = t \wedge Q''_x[t]) \),
			if $x$ is not in \OLA{y}{} and it has another (free) occurrence in 
			$Q'$ or in $Q''$
	\item[(6)]
			\( (\exists \overline{y}) (Q' \wedge x = t \wedge Q'') \)
			replace by
			\( (\exists \overline{y}) (Q'_x[t] \wedge \TRUE \wedge Q''_x[t]) \),
			if $x$ is in \OLA{y}{}
	\item[(7)]
			\( (\exists \overline{y}) (Q' \wedge t = x \wedge Q'') \)
			replace by
			\( (\exists \overline{y}) (Q' \wedge x = t \wedge Q'') \),
			if (i) $t$ is not a variable or (ii) $t$ is a distinct variable from 
			$x$ not occurring in \OLA{y}{} and $x$ is in \OLA{y}{}
\end{enumerate}
The operations for eliminating quantifiers are defined as follows:
\begin{enumerate}
	\item[(8)]
			\( (\exists \OLB{y}{'}) (\exists y) (\exists \OLB{y}{''}) Q \)
			replace by
			\( (\exists \OLB{y}{'}) (\exists \OLB{y}{''}) Q \),
			if $y$ is not free in 
			\( (\exists \OLB{y}{''}) Q \)
	\item[(9)]
			\( (\exists \OLB{y}{_1}) Q_1 \wedge (\exists \OLB{y}{_2}) Q_2 \)
			replace by
			\( (\exists \OLB{z}{_1}) (\exists \OLB{z}{_2}) (R_1 \wedge 
				R_2) \),
			where 
			\( (\exists \OLB{z}{_i}) R_i \)
			is a variant of 
			\( (\exists \OLB{y}{_i}) Q_i \) for $i = 1,2$ such that 
			\( \VARS{R_1} \cap \OLB{z}{_2} = \emptyset \)
			and 
			\( \VARS{R_2} \cap \OLB{z}{_1} = \emptyset \);
			we suppose that 
			\( \OLB{y}{_1} \not = \emptyset \)
			or 
			\( \OLB{y}{_2} \not = \emptyset \) 
\end{enumerate}
Finally, we have:
\begin{enumerate}
	\item[(10)]
			\( A \wedge s = t \)
			replace by
			\( s = t \wedge A \),
			where $A$ is an atom.
	\item[(11)]
			delete any occurrence of the propositional constant \TRUE
	\item[(12)]
			replace $Q$ by \FALSE, if $Q$ obtains at least one occurrence of the 
			propositional constant \FALSE
\end{enumerate}
The algorithm terminates with $Q'$ as the output when no step can be applied to
$Q'$ or when \FALSE\ has been returned. We write \REWRA{Q}{}{Q'} if $Q'$ can be
obtained from $Q$ by one step. By \REWRB{}{} we mean the reflexive and
transitive closure of \REWRA{}{}{}. The following theorem establishes the
correctness and the termination of the solved form algorithm.

\begin{theorem}[\cite{Komara:unif}]\label{cl:Robinson:1}
	The solved form algorithm applied to a query $Q$ will return a query $Q'$
	in solved form after finite number of steps such that
	\TRUTH{\FEA{\LB{}}}{}{Q' \leftrightarrow Q}
	holds.
\end{theorem}

By a {\em rewriting rule\/} \RWA{} we mean any subsets of \REWRB{}{} for
which the {\em uniquess condition\/} (a) and the {\em existence condition\/}
(b) holds. We write \REWRA{Q}{\RWB{}}{Q'} instead of
\( (Q,Q') \in \RWB{} \).
\begin{enumerate}
	\item[(a)]
			if \REWRA{Q}{\RWB{}}{Q'} and \REWRA{Q}{\RWB{}}{Q''}, then 
			\( Q' \equiv Q'' \)
	\item[(b)]
			for any consistent $Q$ there is $Q'$ such that \REWRA{Q}{\RWB{}}{Q'}
\end{enumerate}
Moreover, we suppose that \RWA{} is independent on the choice of names of
variables i.e if $Q_1$ and $Q_2$ are variants and \REWRA{Q_i}{\RWB{}}{Q'_i},
then $Q'_1$ and $Q'_2$ are variants, as well. Note that rewriting rules can be
undefined for inconsistent queries, in particular for the propositional
constant \FALSE.

\paragraph{}
By a (nondeterministic) {\em computation method\/} we mean a pair
\( \CM{} = (\SR{},\RWB{}) \),
where:
\begin{enumerate}
	\item[(a)]
			\SR{} is a {\em selection rule\/}, which selects from any nonempty
			query a nonempty sequence of atoms --- called {\em selected atoms}
	\item[(b)]
			\RWA{} is a rewriting rule
\end{enumerate}
We suppose that the selection rule \SR{} is independent on the choice of names
of variables i.e. if $Q$ and $Q'$ are variants, then selected atoms in $Q'$
directly correspond to selected atoms in $Q$.

Consider a program $P$. We say that $Q'$ is a {\em resolvent\/} of $Q$ and
clauses
\( \OLB{C}{} = C_1, \ldots, C_k \)
from $P$ via \CM{} if $Q'$ can be obtained from $Q$ by performing the following
two steps:
\begin{enumerate}
	\item[(a)]
			First compute a reduction $Q''$ of $Q$ by
			\( C_1, \ldots, C_k \),
			where
			\( A_1, \ldots, A_k \)
			are selected atoms.
	\item[(b)]
			Then \REWRA{Q''}{\RWB{}}{Q'}.
\end{enumerate}
We write \RULEA{Q}{\OLB{C}{}}{Q'} (via \CM{}) or only \RULEA{Q}{P}{Q'} (via
\CM{}). By assumptions on selections and rewriting rules we have that the
notion of resolvent is independent on the choice of names of variables.
Namely, we can state the following claim.

\begin{lemma}\label{cl:oper:variant}
	Let \RULEA{Q_1}{\OLB{C}{}}{Q'_1} via \CM{}. Suppose that $Q_2$ is a variant 
	of $Q_1$. Then there is a (unique) resolvent $Q'_2$ of $Q_2$ and \OLA{C}{}
	via \CM{}. Moreover, $Q'_2$ is a variant of $Q_2$, as well.
\end{lemma}

By a {\em partial SLD-derivation\/} of
\( P \cup \{ Q \} \) 
via \CMB{R} we mean a (possible infinite) sequence
\( Q_0, Q_1, \ldots \)
of queries, where 
\( Q_0 \equiv Q \) 
and $Q_{i+1}$ is a resolvent of $Q_i$ (according some input clauses from the 
program $P$) via \CMB{R}. The derivation will be called {\em SLD-derivation\/}
if it is infinite or the last query $Q_n$ has no resolvent via \CM{}. If the
last query $Q_n$ in this derivation is a consistent \EQ{formula} $E$, then such
derivation will be called {\em SLD-refutation\/} and $E$ a {\em \CMA-computed
answer\/} (or {\em computed answer\/} for short) for
\( P \cup \{ Q \} \).
Finally, we say that this derivation is {\em failed\/} if 
the last query $Q_n$ is an inconsistent \EQ{formula} or a nonempty query having
no resolvent via \CM{}.

\paragraph{}
The soundness of SLD-resolution is based on the following lemma.

\begin{lemma}\label{cl:resolvent}
	Consider a program $P$ and a clause $C$ over a language \LA{}.
	\begin{enumerate}
		\item[(a)]
				If $Q$ is an atomic reduction of an atom $A$ by $C$, then for any 
				model $I$ of \FEA{\LB{}} we have
				\[ \ANS{A}{\TOBB{C}{J}{I}}{J} = \ANS{Q}{I}{J}, \]
				provided $I$ is based on $J$.
		\item[(b)]
				If \RULEA{Q}{P}{Q'}, then
				\( \VARS{Q'} \subseteq \VARS{Q} \)
				and
				\( \TRUTH{P^*}{}{Q' \rightarrow Q} \).
	\end{enumerate}
\end{lemma}

{\bf Proof:}
Let 
\( A \equiv p(s_1, \ldots, s_n) \)
and
\( C' \equiv \CLAUSE{A'}{Q'} \),
where
\( A' \equiv p(t_1, \ldots, t_n) \),
be a clause having no variables in common with $A$ obtained from $C$ by
renaming. Then
\[ Q \equiv (\exists z_1) \ldots (z_m) (s_1 = t_1 \wedge \ldots s_n = t_n
\wedge Q' ) \]
is an atomic reduction of $A$ by $C$, where 
\( \VARS{C'} = \{ z_1, \ldots, z_m \} \).
Consider arbitrary model $I$ for \FEA{\LB{}} based on $J$. Let $h$ be a
solution of $A$ in \TOBA{C'}{J}{I}. Then there is
\( g \in \VV{\LB{}}{J} \)
such that 
\( h(A) \equiv g(A') \)
and \TRUTH{I}{g}{Q'}. Moreover we can suppose that
\( g(x) \equiv h(x) \)
for
\( x \in \VAR \setminus \VARS{C'} \).
Then
\( g(s_i) \equiv h(s_i) \equiv g(t_i) \)
for any $s_i$ and hence \TRUTH{I}{h}{Q}. Assume now that \TRUTH{I}{h}{Q}. Then
there is 
\( g \in \VV{\LB{}}{J} \)
such that 
\( g(x) \equiv h(x) \)
for
\( x \in \VAR \setminus \VARS{C'} \),
\( g(s_i) \equiv g(t_i) \)
for any $s_i$ and \TRUTH{I}{g}{Q'}. We have
\( h(A) \equiv g(A') \)
and hence $h$ is a solution of $A$ in \TOBA{C'}{J}{I}. Since
\( (\forall)C' \)
and
\( (\forall)C \)
are variants, we have
\( \TOA{C'}{J} \equiv \TOA{C}{J} \).
This concludes the proof of (a).

Since Solved Form Algorithm does not introduce new variables, we have that
\( \VARS{Q'} \subseteq \VARS{Q} \),
when $Q'$ is a resolvent of $Q$. Let $I$ based on $J$ be a model for $P^*$.
Then
\( \TOBB{P}{J}{I} \subseteq I \)
by Theorem~\ref{cl:characterization}. If $A$ is a selected atom in $Q$ and
$Q_A$ is an atomic reduction of $A$ by a clause $C$ from $P$, then by (a) we
have
\[ \ANS{Q_A}{I}{J} = \ANS{A}{\TOBB{C}{J}{I}}{J} \subseteq
	\ANS{A}{\TOBB{P}{J}{I}}{J} \subseteq \ANS{A}{I}{J} \]
Consequently
\( \ANS{Q'}{I}{J} \subseteq \ANS{Q}{I}{J} \)
i.e. \TRUTH{I}{}{Q' \rightarrow Q}.
\END

\begin{theorem}[Soundness of SLD-resolution]\label{cl:sound}
	Let $P$ be a program and $Q$ a query. Then any computed answer for 
	\( P \cup \{ Q\} \)
	is a correct answer for 
	\( P \cup \{ Q\} \) 
	as well.
\end{theorem}

{\bf Proof:}
Straightforward by applying previous lemma.
\END

\medskip

Finally we prove the converse of the Soudness Theorem.

\begin{theorem}[Completeness of SLD-resolution]\label{cl:compl}
	Let $P$ be a program, $Q$ a query and \CMB{R} a computation method. If $E$
	is a correct answer for 
	\( P \cup \{ Q \} \), 
	then there is a \CMA-computed answer for 
	\( P \cup \{ Q \} \) 
	more general than $E$.
\end{theorem}

To prove the Completeness Theorem we introduce a modified concept of a value
for queries. Instead of the mapping
\( \ANS{Q}{}{J} : 2^{B_L^J} \rightarrow \VV{L}{J} \),
which has a single interpretation as an argument, we shall consider a mapping 
\( \MANS{Q}{}{J} : 2^{B_L^J} \times \ldots \times 2^{B_L^J}
\rightarrow \VV{L}{J} \)
defined on sequences \OLA{I}{} of interpretations. Each component in \OLA{I}{}
will serve as an ``input'' for the corresponding atom in $Q$. We follow here
ideas developed in \cite{SK:lop}.

By a {\em multiinterpretation\/} based on a pre-interpretation $J$ we mean any
finite sequences of interpretations based on $J$. The empty sequence is denoted
by \EMPTYSEQA. We shall use overlined letters to denote multiinterpretations.
We write
\( \OLB{I}{_1} \OLB{I}{_2} \)
for the concatenation of multiinterpretations \OLA{I}{_1} and \OLA{I}{_2}.
The length of \OLA{I}{} is denoted by \CARD{\OLB{I}{}}. We say that
\OLA{I}{} is for a query $Q$ if
\( \CARD{\OLB{I}{}} = \CARD{Q} \).

Consider a pre-interpretation $J$, a query $Q$ and a multiinterpretation
\OLA{I}{} for $Q$ based on $J$. The value of the query $Q$ in the
multiinterpretation \OLA{I}{} is a set of $J$-valuations, denoted by
\MANS{Q}{\OLB{I}{}}{J}, and it is defined inductively as follows:
\begin{enumerate}
	\item \( \MANS{\TRUE}{\EMPTYSEQB}{J} = \VV{L}{J} \)
			and
			\( \MANS{\FALSE}{\EMPTYSEQB}{J} = \emptyset \)
	\item \( \MANS{s = t}{\EMPTYSEQB}{J} = \SOLN{J}{s = t} \)
	\item \( \MANS{A}{I}{J} = \ANS{A}{I}{J} \),
			if $A$ is an atom.
	\item \( \MANS{Q_1 \wedge Q_2}{\OLB{I}{_1}\OLB{I}{_2}}{J}  = 
				\MANS{Q_1}{\OLB{I}{_1}}{J} \cap \MANS{Q_2}{\OLB{I}{_2}}{J} \),
			where 
			\( \CARD{\OLB{I}{_1}} = \CARD{Q_1} \)
			and 
			\( \CARD{\OLB{I}{_2}} = \CARD{Q_2} \)
	\item \( \MANS{(\exists x)Q}{\OLB{I}{}}{J} = 
				\{ h \in \VV{L}{J} \mid 
					\CCB{h}{x}{d} \in \MANS{Q}{\OLB{I}{}}{J}\ \ 
					{\rm for\ some}\ d \in \UU{L}{J} \} \)
\end{enumerate}
Note that 
\( \ANS{Q}{I}{J} = \MANS{Q}{\OLB{I}{}}{J} \), 
where
\( \OLB{I}{} = I, \ldots, I \)
and
\( \CARD{\OLB{I}{}} = \CARD{Q} \).

In \cite{Komara:unif} the notion of equivalence on queries had been introduced.
We say that queries $Q$ and $Q'$ are {\em equivalent\/} if
\begin{itemize}
	\item $Q$ and $Q'$ have the same number of atoms
	\item for any model $J$ of \FEA{L} and any multiinterpretations \OLA{I}{} 
			based on $J$ (for $Q$) we have
			\( \MANS{Q}{\OLB{I}{}}{J} \equiv \MANS{Q'}{\OLB{I}{}}{J} \)
\end{itemize}

The following theorem states that Solved Form Algorithm preserves equivalence.

\begin{theorem}[\cite{Komara:unif}]\label{cl:Robinson:2}
	Let $Q$ be a query. If \REWRB{Q}{Q'}, then $Q'$ is equivalent to $Q$.
\end{theorem}

Now we restrict our attention only to the Herbrand pre-interpretation \HA\ for
a canonical language \LA{c} of $P$. We shall write
\( \MM{P}{\HC}, \TOA{P}{\HC}, \ldots \)
instead of 
\( \MM{P}{\HB}, \TOA{P}{\HB}, \ldots \).
In the proof of Completeness Theorem we shall deal with multiinterpretations of
the form
\(	\TOCB{P}{}{s_1}, \ldots, \TOCB{P}{}{s_n} \).
We write \TOCA{P}{}{\OLB{s}{}} to denote this multiinterpretation, where
\( \OLB{s}{} = s_1, \ldots, s_n \).
The (finite) sequence \OLA{s}{} of natural numbers will called an {\em
index\/}. We shall use overlined symbols
\( \OLB{s}{}, \OLB{s}{'}, \OLB{s}{''}, \ldots \)
to denote indexes. We write \CARD{\OLB{s}{}} for the length of the index
\OLA{s}{}. The empty sequence is denoted by \EMPTYINDA. We apply the multiset
ordering $\preceq$ described in \cite{DM:multiset} to indexes. Let \OLA{s}{'}
and \OLA{t}{'} be indexes obtained from \OLA{s}{} and \OLA{t}{} by sorting them
in decreasing order. We write
\( \OLB{s}{} \prec \OLB{t}{} \)
if \OLA{s}{'} precedes \OLA{t}{'} lexicogrphically. The ordering has Noetherian
property i.e. there is no infinite decreasing sequence
\( \OLB{s}{_0} \succ \OLB{s}{_1} \succ \OLB{s}{_2} \succ \ldots \).

Given a program $P$ we assign to any \EQ{formula} $E$ and a query $Q$ a set of
indexes, denoted by \IND{P}{E}{Q}, as follows:
\[ \OLB{s}{} \in \IND{P}{E}{Q}\ \ {\rm iff}\ \ 
	\SOLN{\HC}{E} \subseteq \MANS{Q}{\TOCB{P}{\HC}{\OLB{s}{}}}{\HC} \]
Note that if $Q$ is an empty query $E'$ then 
\( \IND{P}{E}{E'} \not = \emptyset \)
iff $E'$ is more general than $E$ by Proposition~\ref{cl:order}.

\begin{proposition}[Soundness of Index Set]\label{cl:index:sound}
	Consider a program $P$, a query $Q$ and an \EQ{formula} $E$. Then
	\begin{equation}\label{eq:sound:index}
		\IND{P}{E}{Q} \not = \emptyset\ \ {\rm iff}\ \ \TRUTH{P^*}{}{E 
		\rightarrow Q}\ {\rm in}\ \LB{P}
	\end{equation}
\end{proposition}

{\bf Proof:}
Using Theorem~\ref{cl:E:canonical} by noting that \IND{P}{E}{Q} is nonempty if
only if \TRUTH{\TOCB{P}{\HC}{n}}{}{E \rightarrow Q} for some $n$.
\END

\medskip

\begin{proposition}[Local Step]\label{cl:index:local}
	Consider a program $P$ and a computation method \CMB{R}. Let $Q$ be a 
	nonempty query and 
	\( \OLB{s}{} \in \IND{P}{E}{Q} \).
	Then there is an index \OLA{s}{'} and a query $Q'$ such that
	\begin{enumerate}
		\item \RULEA{Q}{P}{Q'} via \CMB{R}
		\item \( \OLB{s}{'} \in \IND{P}{E}{Q'} \)
		\item \( \OLB{s}{'} \prec \OLB{s}{} \)
	\end{enumerate}
\end{proposition}

{\bf Proof:}
The proof is contained in the Appendix.
\END

\medskip

Now we are ready to prove the Completeness Theorem.

\medskip

{\bf Proof of Theorem~\ref{cl:compl}:}
Suppose that $E$ is a correct
answer for 
\( P \cup \{ Q \} \). 
Then \IND{P}{E}{Q} is nonempty and hence there is a derivation 
\( Q_0, Q_1, Q_2, \ldots \)
for 
\( P \cup \{ Q \} \) 
via \CMB{R} and a sequence of indexes 
\( \overline{s}_0 \succ \overline{s}_1 \succ \overline{s}_2 \succ \ldots \)
such that 
\( \overline{s}_i \in \IND{P}{E}{Q_i} \) 
for any $i$. The sequence of indexes must be finite and hence the derivation
has a finite length, say $n$. Then the query $Q_n$ is consistent, since
\IND{P}{E}{Q_n} is nonempty. Moreover $Q_n$ must be a (consistent)
\EQ{formula} $E'$.  Thus the derivation is a refutation. Then for the computed
answer $E'$ of this refutation we have
\( \IND{P}{E}{E'} \not = \emptyset \). 
So 
\TRUTH{P^*}{}{E \rightarrow E'} 
and hence 
\TRUTH{\FEA{L}}{}{E \rightarrow E'}.
Thus $E'$ is more general than $E$. This concludes the proof.
\END

\paragraph{}
Now, we make some comments about our approach and compare it to standard ones.
\begin{enumerate}
	\item By allowing of equations and existentially quantified variables in 
			queries we avoid problems when dealing with new variables. As a 
			consequence of Lemma~\ref{cl:oper:variant} we have that the 
			operational semantics is independent on the choice of names of new 
			variables introducing by input clauses. Really, let
			\( Q_0, Q_1, Q_2, \ldots \)
			be an SLD-derivation of
			\( P \cup \{ Q \} \),
			where $Q_{i+1}$ is a resolvent of $Q_i$ and \OLA{C}{_{i}} via \CMB{R}.
			If $Q'$ is a variant of $Q$, then there is an SLD-derivation 
			\( Q'_0, Q'_1, Q'_2, \ldots \)
			of
			\( P \cup \{ Q \} \)
			such that  $Q'_{i+1}$ is a resolvent of $Q'_i$ and \OLA{C}{_{i}} via 
			\CMB{R}. Moreover every $Q'_i$ is a variant of $Q_i$.
	\item Secondly, the notion of
			computed answer in our approach has nice transitive properties. 
			Namely, if $Q'$ is a resolvent of $Q$ and $E$ is a computed answer for
			$Q'$, then $E$ is a computed answer for $Q$, as well. This is clearly
			true, since we may concatenate two partial derivations directly.
	\item Thirdly, our approach can be viewed as a basis for concurrent logic 
			languages, since the notion of parallel composition can be defined
			directly --- as a consistent conjunction of \EQ{formulas}. Consider 
			again the queries
			\( Q_1 \equiv p(x) \) 
			and
			\( Q_2 \equiv q(y) \),
			and the program 
			\( \{ \CLAUSE{p(f(z))}{}, \CLAUSE{q(g(z))}{} \} \).
			If we run $Q_1$ and $Q_2$ parallel
			and independently, we could get answers
			\( E_1 \equiv (\exists z) (x = f(z)) \) 
			for $Q_1$ and
			\( E_2 \equiv (\exists z) (x = g(z)) \) 
			for $Q_2$. By the parallel composition of $E_1$ and $E_2$ we obtain an
			answer
			\[ (\exists z) (x = f(z)) \wedge (\exists z) (y = g(z)) \]
			for 
			\( Q \equiv Q_1 \wedge Q_2 \),
			which is equivalent to 
			\[ (\exists z_1) (\exists z_2) (x = f(z_1) \wedge y = g(z_2)) \]
			The corresponding finite substitution is
			\[ \sigma = \{ \BINDA{x}{f(z_1)}, \BINDA{y}{g(z_2)} \} \]
\end{enumerate}

\appendix
\section{Appendix: Remaining Proofs}\label{sc:app}

We first introduce the notion of goodness for interpretations 
(see \cite{Apt:logic}).

\begin{definition}\label{df:good}
	Let $J$ be a pre-interpretation which is a model of \FEA{L}. We say that an 
	interpretation $I$ based on $J$ is {\em good\/} iff for every atom $A$ there 
	are \EQ{formulas}
	\( E_1, \ldots, E_n \)
	such that
	\[ \TRUTH{I}{}{(E_1 \vee \ldots \vee E_n) \leftrightarrow A} \]
\end{definition}

Since \TRUTH{\BB{L}{J}}{}{\TRUE \leftrightarrow A} and
\TRUTH{\emptyset}{}{\FALSE \leftrightarrow A}, we have that \BB{L}{J} and 
$\emptyset$ are good interpretations.

\begin{lemma}\label{cl:good:1}
	Let $J$ be a pre-interpretation which is a model of \FEA{L}. The 
	interpretation based on $J$ is good iff for each query $Q$ there are 
	\EQ{formulas} 
	\( E_1, \ldots, E_n \)
	such that
	\[ \TRUTH{I}{}{(E_1 \vee \ldots \vee E_n) \leftrightarrow Q} \]
\end{lemma}

{\bf Proof:}
Straightforward by structural induction since (i)
\TRUTH{I}{}{(E_1 \vee \ldots \vee E_n) \leftrightarrow Q}
and
\TRUTH{I}{}{(E'_1 \vee \ldots \vee E'_m) \leftrightarrow Q'}
implies that
\TRUTH{I}{}{ \bigvee_{i,j} (E_i \wedge E'_j) \leftrightarrow (Q \wedge Q')}
holds and (ii)
\TRUTH{I}{}{(E_1 \vee \ldots \vee E_n) \leftrightarrow Q}
implies
\TRUTH{I}{}{((\exists x)E_1 \vee \ldots \vee (\exists x)E_n) \leftrightarrow 
(\exists x)Q}.
\END

\medskip

\begin{lemma}\label{cl:good:2}
	Let $J$ be a pre-interpretation which is a model of \FEA{L} and $P$ a 
	program over \LA{}. Let $I$ be an interpretation based on $J$. Suppose $I$ 
	is good. Then \TOBA{P}{J}{I} is good, as well.
\end{lemma}

{\bf Proof:}
Consider an atom 
\( A \equiv p(s_1, \ldots, s_n) \).
The operator \TOA{P}{J} does not depend on the choice of
the names of variables in $P$. Thus we can assume that each clause has no
variables in common with $A$.

Suppose now that 
\( C \equiv \CLAUSE{p(t_1, \ldots, t_n)}{Q} \) 
is a clause from $P$ containing variables
\( \OLB{z}{} = z_1, \ldots, z_m \).
Then we have (see  Lemma~\ref{cl:resolvent})
\[	\ANS{A}{\TOBB{C}{J}{I}}{J} = 
	\ANS{(\exists \OLB{z}{})(s_1 = t_1 \wedge \ldots \wedge s_n = t_n 
			\wedge Q)}{I}{J} \]
But $I$ is good, so by Lemma~\ref{cl:good:1} we have
\TRUTH{I}{}{(E'_1 \vee \ldots \vee E'_k) \leftrightarrow Q}
for some \EQ{formulas}
\( E'_1, \ldots, E'_k \).
Put
\( E_i \equiv (\exists \OLB{z}{}) (s_1 = t_1 \wedge \ldots \wedge s_n = t_n 
\wedge E'_i) \).
Then
\( \ANS{A}{\TOBB{C}{J}{I}}{J} = \SOLN{J}{E_1 \vee \ldots \vee E_k} \).
Consequently for any clause $C$ there is a disjunction $D_C$ of \EQ{formulas}
such that
\( \ANS{A}{\TOBB{C}{J}{I}}{J} = \SOLN{J}{D_C} \).
Hence
\[ \TRUTH{\TOBB{P}{J}{I}}{}{(D_{C_1} \vee \ldots \vee D_{C_l}) \leftrightarrow 
A}, \]
where
\( P = \{ C_1, \ldots, C_l \} \).
\END

\medskip

By a straightforward refinement of the proofs of previous lemmas we have the
following claim.

\begin{lemma}\label{cl:good:3}
	Let $J$ be a pre-interpretation which is a model of \FEA{\LA{}} and $P$ a
	program over \LA{}. Then for every 
	\( n \geq 0 \)
	\TOCA{P}{J}{n} is good. Moreover, for each query $Q$ we can find 
	\EQ{formulas}
	\( E_1, \ldots, E_m \)
	containing the only function symbols from $P$ and $Q$ such that
	\[ \TRUTH{\TOCB{P}{J}{n}}{}{(E_1 \vee \ldots \vee E_m) \leftrightarrow Q} \]
\end{lemma}

Let $F$, $F'$ be equational formulas and $J$ a pre-interpretation. We write
\( F \preceq_J F' \)
or
\( F \approx_J F' \)
if
\( \SOLN{J}{F} \subseteq \SOLN{J}{F'} \)
or
\( \SOLN{J}{F} = \SOLN{J}{F'} \),
respectively.

\begin{proposition}[\cite{Komara:unif}]\label{cl:compactness}
	Suppose that $L$ contains infinitely many constants and \HA\ is the 
	Herbrand pre-interpretation for $L$. Let $E$ and $E_1$, \ldots, $E_n$ be 
	\EQ{formulas} over $L$.
	\begin{enumerate}
		\item If 
				\( E_1 \prec_{\HB} E , \ldots, E_n \prec_{\HB} E \), 
				then 
				\( E_1 \vee \ldots \vee E_n \prec_{\HB} E \).
		\item If 
				\( E \approx_{\HB} E_1 \vee \ldots \vee E_n \), 
				then 
				\( E \approx_{\HB} E_j \) 
				for some $E_j$.
		\item If 
				\( E \preceq_{\HB} E_1 \vee \ldots \vee E_n \), 
				then 
				\( E \preceq_{\HB} E_j \) 
				for some $E_j$.
	\end{enumerate}
\end{proposition}

\subsection{Proof of Proposition~\protect{\ref{cl:F:canonical}}}\label{sc:app:canonical}

We shall omit the supperscript \HA\ in \MM{P}{\HB}, \TOA{P}{\HB}, \ldots.

\medskip

{\bf Proof of Proposition~\ref{cl:F:canonical}}\\
\( (a) \Rightarrow (b) \)\\
Straightforward.\\
\( (b) \Rightarrow (c) \)\\
Straightforward by Lemma~\ref{cl:extension}. \\
\( (d) \Rightarrow (c) \)\\
Straightforward by noting that
\( \TOCB{P}{\HC}{n} \subseteq \MM{P}{\HC} \)
and since $F$ is positive.
\END

\medskip

To prove that (c) implies (a) we first show that the following lemma holds. 

\begin{lemma}\label{cl:F:canonical:1}
	Let $F$ be a closed positive formula over \LA{c}. If
	\TRUTH{\MM{P}{\HC}}{}{F}, then \TRUTH{P}{}{F} in \LA{c}.
\end{lemma}

{\bf Proof:}
If
\TRUTH{\MM{P}{\HC}}{}{s = t},
then 
\( s \equiv t \)
and hence \TRUTH{P}{}{s = t} in \LA{c}. Consider an atom $A$. Then by
Theorem~\ref{cl:characterization} \TRUTH{\MM{P}{\HC}}{}{A} implies
\TRUTH{P}{}{A} in \LA{c}. By a straightforward application of the induction
hypothesis we can prove the claim for a formula $F$ of the form
\( F \equiv G \wedge H \) 
or
\( F \equiv G \vee H \).

Suppose 
\( F \equiv (\exists x) G \). 
If 
\( \TRUTH{\MM{P}{\HC}}{}{F} \), 
then 
\( \TRUTH{\MM{P}{\HC}}{}{G_x[s]} \) 
for some ground term $s$. By induction hypothesis we have 
\( \TRUTH{P}{}{G_x[s]} \) 
in \LA{c}. Since 
\( G_x[s] \rightarrow (\exists x) G \) 
is valid, we have 
\( \TRUTH{P}{}{F} \)
in \LA{c}.

Assume finally that 
\( F \equiv (\forall x) G \). 
If 
\( \TRUTH{\MM{P}{\HC}}{}{F} \), 
then 
\( \TRUTH{\MM{P}{\HC}}{}{G_x[s]} \)  
for any ground term $s$. By the induction hypothesis we have 
\( \TRUTH{P}{}{G_x[s]} \) 
in \LA{c}. If $s$ is a constant $a$ not occurring in $P$ and $G$, then by
Theorem~\ref{cl:constants} we have 
\( \TRUTH{P}{}{G} \) 
in \LA{c} and therefore
\( \TRUTH{\TOCB{P}{\HC}{n}}{}{F} \)
in \LA{c}.
\END

\medskip
{\bf Proof of Proposition~\ref{cl:F:canonical} continued:}\\
\( (c) \Rightarrow (a) \)\\
Let $F$ be arbitrary (not necessary closed) positive formula over \LA{P}.
Consider distinct constants
\( a_1, \ldots, a_k \) 
from \LA{c} not occurring in $P$ and $F$. If \TRUTH{\MM{P}{\HC}}{}{F}, then 
\TRUTH{\MM{P}{\HC}}{}{F[a_1, \ldots, a_k]} and hence from the lemma above 
we have \TRUTH{P}{}{F[a_1, \ldots, a_k]} in \LA{c}. Theorem~\ref{cl:constants}
implies that \TRUTH{P}{}{F} in \LA{c} and hence \TRUTH{P}{}{F} in \LA{P}.
\END

\medskip

To prove that (c) implies (d) we need the following claim. 

\begin{lemma}\label{cl:F:canonical:2}
	Let $F$ be a positive formula over \LA{c}. If
	\( a_1, \ldots, a_k \) 
	are distinct constants from \LA{c} not occurring in $P$ and $F$, then:
	\begin{equation}\label{eq:F:canonical:2}
	\TRUTH{\TOCB{P}{\HC}{n}}{}{F[a_1, \ldots, a_k]}\ \ {\rm iff}\ \ 
	\TRUTH{\TOCB{P}{\HC}{n}}{}{F}
	\end{equation}
\end{lemma}

We will first show how 
Proposition~\ref{cl:F:canonical} \( (c) \Rightarrow (d) \)
can be derived from Lemma~\ref{cl:F:canonical:2}.

\medskip

{\bf Proof of Proposition~\ref{cl:F:canonical} continued:}\\
\( (c) \Rightarrow (d) \)\\
First we prove by structural induction that the claim holds for closed positive
formulas over \LA{c} using analogous arguments as those in the proof of
Lemma~\ref{cl:F:canonical:1}. The only interesting cases are that if $F$ is an
atom $A$ or it is of the form
\( (\forall x) G \). 
Consider an atom $A$. Then by Theorem~\ref{cl:characterization}
\TRUTH{\MM{P}{\HC}}{}{A} implies
\( A \in \TOCB{P}{\HC}{n} \)
for some $n$. Let 
\( \TRUTH{\MM{P}{\HC}}{}{(\forall x) G} \).
Then 
\( \TRUTH{\MM{P}{\HC}}{}{G_x[s]} \)  
for any ground term $s$. By the induction hypothesis we have 
\( \TRUTH{\TOCB{P}{\HC}{n}}{}{G_x[s]} \) 
for some $n$. If $s$ is a constant $a$ not occurring in $P$ and $G$, then by
Lemma~\ref{cl:F:canonical:2} we have 
\( \TRUTH{\TOCB{P}{\HC}{n}}{}{G} \) 
and therefore
\( \TRUTH{\TOCB{P}{\HC}{n}}{}{(\forall x) G} \).

Now let $F$ be arbitrary (not necessary closed) positive formula over \LA{P}.
Consider distinct constants
\( a_1, \ldots, a_k \) 
from \LA{c} not occurring in $P$ and $F$. If \TRUTH{\MM{P}{\HC}}{}{F}, then 
\TRUTH{\MM{P}{\HC}}{}{F[a_1, \ldots, a_k]} and hence from the proof above 
we have 
\TRUTH{\TOCB{P}{\HC}{n}}{}{F[a_1, \ldots, a_k]} for some $n$.
Lemma~\ref{cl:F:canonical:2} implies that
\TRUTH{\TOCB{P}{\HC}{n}}{}{F}.
This concludes the proof.
\END

\medskip

The proof of Lemma~\ref{cl:F:canonical:2} is based on Theorem on Constants and
the next claim.

\begin{lemma}\label{cl:F:canonical:3}
	For every 
	\( n \geq 1 \)
	there is a program $P_n$ over \LA{P} containing only facts such that
	\( \MM{P_n}{\HC} = \TOCB{P}{\HC}{n} \).
\end{lemma}

{\bf Proof:}
Observe first that if $P'$ contains only facts then \MM{P'}{\HC} is good, 
since 
\( \MM{P'}{\HC} = \TOCB{P'}{\HC}{1} \).
We prove the claim by induction on $n$. Suppose first $n=1$. Clearly, if 
$P_1$ contains only facts from $P$ having consistent bodies, then
\( \MM{P_1}{\HC} = \TOCB{P}{\HC}{1} \).

Now suppose that the result holds for 
\( n \geq 1 \).
Then by the induction hypothesis there is a program $P_n$ over \LA{P} 
containing only facts such that
\( \MM{P_n}{\HC} = \TOCB{P}{\HC}{n} \).
Hence
\( \TOCB{P}{\HC}{(n+1)} = \TOBB{P}{\HC}{\MM{P_n}{\HC}} \).
Let 
\( C \equiv \CLAUSE{A}{Q} \)
be a clause from $P$. Since \MM{P_n}{\HC} is good, then by 
Lemma~\ref{cl:good:3} there are \EQ{formulas}
\( E_1, \ldots, E_m \)
over \LA{P} such that
\( \TRUTH{\MM{P_n}{\HC}}{}{(E_1 \vee \ldots \vee E_k) \leftrightarrow Q} \).
Put 
\( P_C = \bigcup_{i=1}^{k} \{ \CLAUSE{A}{E_i} \mid E_i\ {\rm is\ consistent} 
\} \) 
and
\( P_{n+1} = \bigcup_{C \in P} P_C \).
Clearly $P_{n+1}$ is over \LA{P}.

Let 
\( p(d_1, \ldots, d_m) \in \TOBB{P}{\HC}{\MM{P_n}{\HC}} \).
Then there is a clause 
\( C \equiv \CLAUSE{A}{Q} \)
from $P$ and a valuation $h$ such that
\( p(d_1, \ldots, d_m) \equiv h(A) \)
and \TRUTH{\MM{P_n}{\HC}}{h}{Q}. Then for some
\( \CLAUSE{A}{E} \in P_C \)
we have \TRUTH{\MM{P_n}{\HC}}{h}{E} and hence
\TRUTH{\TOCB{P_{n+1}}{\HC}{0}}{h}{E}.
So \( p(d_1, \ldots, d_m) \in \MM{P_{n+1}}{\HC} \).

Now let
\( p(d_1, \ldots, d_m) \in \MM{P_{n+1}}{\HC} \).
Then there is a fact
\CLAUSE{A}{E}
from $P_C$, where
\( C \equiv \CLAUSE{A}{Q} \),
and a valuation $h$ such that
\( p(d_1, \ldots, d_m) \equiv h(A) \)
and \TRUTH{\MM{P_{n+1}}{\HC}}{h}{E}. 
Then we have \TRUTH{\MM{P_n}{\HC}}{h}{Q} and hence
\( p(d_1, \ldots, d_m) \in \TOBB{P}{\HC}{\MM{P_n}{\HC}} \).
\END

\medskip

{\bf Proof of Lemma~\ref{cl:F:canonical:2}:}
Consider a program $P_n$ from Lemma~\ref{cl:F:canonical:3}. Let \LA{P_n}
be a language obtained from \LA{P} by adding constants
\( a_1, \ldots, a_k \)
and symbols from $F$. So the program $P_n$ is over \LA{P_n} and \LA{c}
is the canonical language for $P_n$. If
\( \MM{P_n}{\HC} = \TRUTH{\TOCB{P}{\HC}{n}}{}{F[a_1, \ldots, a_k]} \),
then by the implication
\( (c) \Rightarrow (a) \) 
in Proposition~\ref{cl:F:canonical} we obtain
\TRUTH{P_n}{}{F[a_1, \ldots, a_k]} in \LA{P_n}. So by
Theorem~\ref{cl:constants}
\TRUTH{P_n}{}{F} in \LA{P_n} and hence 
\( \TOCB{P}{\HC}{n} = \TRUTH{\MM{P_n}{\HC}}{}{F} \).
\END

\subsection{Proof of Proposition~\protect{\ref{cl:index:local}}}\label{sc:app:comp}

We first generalize the notion of goodness for multiinterpretations.

\begin{definition}\label{df:strong:good}
	Let $J$ be a pre-interpretation which is a model of \FEA{\LB{}}. We say that 
	a multiinterpretation
	\( \OLB{I}{} = I_1, \ldots, I_n \)
	based on $J$ is good if each $I_i$ is a good interpretation.
\end{definition}

By the definition the empty multiinterpretation \EMPTYSEQA\ is good. In the
sequel we shall write
\TOBA{P}{J}{\OLB{I}{}} instead of
\( \TOBB{P}{J}{I_1}, \ldots, \TOBB{P}{J}{I_n} \).
As a straightforward corollary of Lemma~\ref{cl:good:2} we have the following
claim.

\begin{lemma}\label{cl:strong:good:1}
	Let $J$ be a pre-interpretation which is a model of \FEA{\LB{}} and $P$ a 
	program over \LA{}. Let \OLA{I}{} be a multiinterpretation based on $J$. 
	Suppose \OLA{I}{} is good. Then \TOBA{P}{J}{\OLB{I}{}} is good, as well.
\end{lemma}

In the next lemmas we shall use the following notations. Let $F$ be an
equational formula and \OLA{I}{} a multiinterpretation based on $J$ for a query
$Q$. We write
\( \MTRUTH{\OLB{I}{}}{}{F \leftrightarrow Q} \)
or
\( \MTRUTH{\OLB{I}{}}{}{F \rightarrow Q} \)
instead of 
\( \SOLN{J}{F} = \MANS{Q}{\OLB{I}{}}{J} \)
or
\( \SOLN{J}{F} \subseteq \MANS{Q}{\OLB{I}{}}{J} \),
respectively.

\begin{lemma}\label{cl:strong:good:2}
	Let $J$ be a pre-interpretation which is a model of \FEA{L}. A 
	multiinterpretation \OLA{I}{} based on $J$ is good iff for each query
	$Q$, where
	\( \CARD{Q} = \CARD{\OLB{I}{}} \),
	there are \EQ{formulas} 
	\( E_1, \ldots, E_n \)
	such that
	\[ \MTRUTH{\OLB{I}{}}{}{(E_1 \vee \ldots \vee E_n) \leftrightarrow Q} \]
\end{lemma}

{\bf Proof:}
Straightforward by induction since (i)
\( \MTRUTH{\OLB{I}{_1}}{}{(E_1 \vee \ldots \vee E_n) \leftrightarrow Q} \)
and
\( \MTRUTH{\OLB{I}{_2}}{}{(E'_1 \vee \ldots \vee E'_m) \leftrightarrow Q'} \)
implies that
\( \MTRUTH{\OLB{I}{_1}\OLB{I}{_2}}{}{ \bigvee_{i,j} (E_i \wedge E'_j) 
\leftrightarrow (Q \wedge Q')} \)
holds and (ii) 
\( \MTRUTH{\OLB{I}{}}{}{(E_1 \vee \ldots \vee E_n) \leftrightarrow Q} \)
implies
\( \MTRUTH{\OLB{I}{}}{}{((\exists x)E_1 \vee \ldots \vee (\exists x)E_n) 
\leftrightarrow (\exists x)Q} \).
\END

\medskip

Now consider arbitrary but fixed program $P$. Let \LA{c} be a canonical
language for $P$ and \HA\ a Herbrand pre-interpretation for \LA{c}. 
From now we omit the supperscript \HA\ in \MM{P}{\HB}, \TOA{P}{\HB}, \ldots. If
\( \OLB{C}{} = C_1, \ldots C_n \) 
is a sequence of clauses and 
\( \OLB{I}{} = I_1, \ldots I_n \) 
is a multiinterpretation based on \HA\ we write 
\TOBA{\OLB{C}{}}{\HC}{\OLB{I}{}} 
instead of 
\( \TOBB{C_1}{}{I_1}, \ldots, \TOBB{C_n}{}{I_n} \). 
Using the similar arguments as in the proof above we can show that the
following holds:
\begin{equation}\label{eq:app:index:1}
	\MANS{Q}{\TOBB{P}{\HC}{\OLB{I}{}}}{\HC} = 
	\bigcup_{\OLB{C}{}} \MANS{Q}{\TOBB{\OLB{C}{}}{\HC}{\OLB{I}{}}}{\HC},
\end{equation}
where the union on the right hand side is by all possible sequences of clauses
from $P$. The following lemma plays a crucial role in the proof of
Proposition~\ref{cl:index:local}.

\begin{lemma}\label{cl:strong:good:3}
	Consider a multiinterpretation \OLA{I}{} for a query $Q$ and an \EQ{formula}
	$E$. Suppose that \OLA{I}{} is good and
	\( \MTRUTH{\TOBB{P}{\HC}{\OLB{I}{}}}{}{E \rightarrow Q} \).
	Then there is a sequence \OLA{C}{} of clauses from $P$ such that
	\( \MTRUTH{\TOBB{\OLB{C}{}}{\HC}{\OLB{I}{}}}{}{E \rightarrow Q} \).
\end{lemma}

{\bf Proof:}
Consider arbitrary sequence \OLA{C}{} of clauses from $P$, where
\( \CARD{\OLB{C}{}} = \CARD{Q} \).
By Lemma~\ref{cl:strong:good:1} \TOBA{\OLB{C}{}}{\HC}{\OLB{I}{}} is good, as
well. Then there is a disjunction $D_{\OLB{C}{}}$ of \EQ{formulas}
such that
\MTRUTH{\TOBB{\OLB{C}{}}{\HC}{\OLB{I}{}}}{}{D_{\OLB{C}{}}
\leftrightarrow Q}.
By (\ref{eq:app:index:1}) we have
\MTRUTH{\TOBB{P}{\HC}{\OLB{I}{}}}{}{ (D_{\OLB{C}{_1}} \vee
\ldots \vee D_{\OLB{C}{_k}}) \leftrightarrow Q},
where 
\( \OLB{C}{_1}, \ldots, \OLB{C}{_k} \)
are all possible sequences of clauses from $P$ with the length \CARD{Q}.
Suppose now that 
\( \MTRUTH{\TOBB{P}{\HC}{\OLB{I}{}}}{}{E \rightarrow Q} \)
holds.
Then by Theorem~\ref{cl:compactness} there is \OLA{C}{} such that
\( \SOLN{\HC}{E} \subseteq \SOLN{\HC}{D_{\OLB{C}{}}} \).
This implies 
\( \MTRUTH{\TOBB{\OLB{C}{}}{\HC}{\OLB{I}{}}}{}{E \rightarrow Q} \).
\END

\medskip

Now we are in position to prove the desired proposition.

\medskip

{\bf Proof of Proposition~\ref{cl:index:local}:}
Let
\( \OLB{s}{} = s_1, \ldots, s_n \in \IND{P}{E}{Q} \),
where $Q$ is nonempty. Note that
\( s_i > 0 \)
for each $s_i$. Put 
\( \OLB{s}{''} = s''_1, \ldots, s''_n \), 
where
\( s''_i = s_i - 1 \).
Then
\( \MTRUTH{\TOBB{P}{\HC}{\TOCB{P}{\HC}{\OLB{s}{''}}}}{}{E \rightarrow Q} \)
and hence by Lemma~\ref{cl:strong:good:3} there is a sequence 
\( \OLB{C}{} = C_1, \ldots, C_n \)
of clauses from the program $P$ such that
\( \MTRUTH{\TOBB{\OLB{C}{}}{\HC}{\TOCB{P}{\HC}{\OLB{s}{''}}}}{}{E \rightarrow 
Q} \).
We define an index
\( \OLB{s}{'} = \OLB{s_1}{'}, \ldots, \OLB{s_n}'{} \)
and a sequence 
\( Q_1, \ldots, Q_n \) 
of queries as follows:
\begin{enumerate}
	\item Assume that the $i$-th atom $A_i$ in $Q$ is selected. Then we define 
			$Q_i$ as the atomic reduction of $A_i$ by $C_i$ and 
			\[ \OLB{s_i}{'} = \overbrace{s''_i, \ldots, s''_i}^{k - times}, \]
			where $k$ is the number of atoms in the body of $C_i$.
			Then by Lemma~\ref{cl:resolvent} we have
			\begin{equation}\label{eq:app:index:2}
				\MANS{A_i}{\TOBB{C_i}{\HC}{\TOCB{P}{\HC}{s''_i}}}{\HC} = 
				\ANS{A_i}{\TOBB{C_i}{\HC}{\TOCB{P}{\HC}{s''_i}}}{\HC} = 
				\ANS{Q_i}{\TOCB{P}{\HC}{s''_i}}{\HC} =
				\MANS{Q_i}{\TOCB{P}{\HC}{\OLB{s_i}{'}}}{\HC}
			\end{equation}
	\item If the $i$-th atom $A_i$ in $Q$ is not selected, we put 
			\( Q_i \equiv A_i \)
			and 
			\( \OLB{s_i}{'} = s_i \).
			Then
			\begin{equation}\label{eq:app:index:3}
				\MANS{A_i}{\TOBB{C_i}{\HC}{\TOCB{P}{\HC}{s''_i}}}{\HC} 
				\subseteq
				\MANS{Q_i}{\TOCB{P}{\HC}{\OLB{s_i}{'}}}{\HC}
			\end{equation}
\end{enumerate}
Clearly
\( \OLB{s}{'} \prec \OLB{s}{} \).
Let $Q''$ be obtained from $Q$ by simultaneously replacing of each $A_i$ by
$Q_i$. So $Q''$ is the reduction of $Q$ by \OLA{C}{}. By (\ref{eq:app:index:2})
and (\ref{eq:app:index:3}) we have
\[
	\MANS{Q}{\TOBB{\OLB{C}{}}{\HC}{\TOCB{P}{\HC}{\OLB{s}{''}}}}{\HC} \subseteq
	\MANS{Q''}{\TOCB{P}{\HC}{\OLB{s}{'}}}{\HC}
\]
Then 
\( \OLB{s}{'} \in \IND{P}{E}{Q''} \)
and hence $Q''$ is consistent. Therefore there is a resolvent $Q'$ of $Q$ via
\CMB{R} equivalent to $Q''$. By Theorem~\ref{cl:Robinson:2} we have
\[
	\MANS{Q}{\TOBB{\OLB{C}{}}{\HC}{\TOCB{P}{\HC}{\OLB{s}{''}}}}{\HC} \subseteq
	\MANS{Q'}{\TOCB{P}{\HC}{\OLB{s}{'}}}{\HC}
\]
and hence
\( \OLB{s}{'} \in \IND{P}{E}{Q'} \).
This concludes the proof.
\END

\end{document}